\documentclass[twocolumn,eqsecnum,showpacs,showkeys,amsmath,amssymb,nofootinbib,superscriptaddress,floatfix]{revtex4}

\usepackage{graphicx}
\usepackage{subfigure}
\usepackage{bm}

\def\vec#1{\mathchoice{\mbox{\boldmath$\displaystyle#1$}}
{\mbox{\boldmath$\textstyle#1$}}
{\mbox{\boldmath$\scriptstyle#1$}}
{\mbox{\boldmath$\scriptscriptstyle#1$}}}
\makeatletter
\newcommand\erfc{\mathop{\operator@font erfc}\nolimits}
\def\slashchar#1{\setbox0=\hbox{$#1$}
   \dimen0=\wd0 \setbox1=\hbox{/} \dimen1=\wd1
   \ifdim\dimen0>\dimen1 \rlap{\hbox to \dimen0{\hfil/\hfil}} #1
   \else  \rlap{\hbox to \dimen1{\hfil$#1$\hfil}} / \fi}

\makeatother

\begin{document}
 
\title{Fluctuating initial conditions in 
heavy-ion collisions from the Glauber approach\footnote{Supported by 
Polish Ministry of Science and Higher Education under
grant N202~034~32/0918}}

\author{Wojciech Broniowski} 
\email{Wojciech.Broniowski@ifj.edu.pl} 
\affiliation{Institute of Physics, \'Swi\c{e}tokrzyska Academy,
ul.~\'Swi\c{e}tokrzyska 15, PL-25406~Kielce, Poland} 
\affiliation{The H. Niewodnicza\'nski Institute of Nuclear Physics,
PL-31342 Krak\'ow, Poland}
\author{Piotr Bo\.zek}
\email{Piotr.Bozek@ifj.edu.pl}
\affiliation{Institute of Physics, Rzesz\'ow University, PL-35959 Rzesz\'ow, Poland}
\affiliation{The H. Niewodnicza\'nski Institute of Nuclear Physics,
PL-31342 Krak\'ow, Poland}
\author{Maciej Rybczy\'nski}
\email{Maciej.Rybczynski@pu.kielce.pl}
\affiliation{Institute of Physics, \'Swi\c{e}tokrzyska Academy,
ul.~\'Swi\c{e}tokrzyska 15, PL-25406~Kielce, Poland}
\date{28 June}

\begin{abstract}
In the framework of the Glauber approach applied to the initial stage 
of ultra-relativistic heavy-ion collisions we analyze the shape 
parameters of the early-formed system (fireball) and their event-by-event fluctuations.
We test a variety of models: the conventional wounded 
nucleon model, a model admixing binary collisions to the wounded nucleons, 
a model with {\em hot spots}, as well as 
the hot-spot model where the deposition of energy occurs with a superimposed probability distribution.  
We look in detail at the so-called {\em participant} harmonic moments, $\varepsilon^\ast$, obtained by an averaging procedure where 
in each event the system is translated to its center of mass and aligned with the major principal axis of the ellipse of inertia.
Quantitative comparisons indicate substantial 
relative effects for $\varepsilon^\ast$ in variants of Glauber models. 
On the other hand, the dependence of the scaled 
standard deviation $\Delta \varepsilon^\ast/\varepsilon^\ast$ 
on the chosen model is weak. 
For all models the values range from about 0.5 for the central collisions to about 0.3-0.4 for peripheral collisions, both for 
the gold-gold and copper-copper collisions. They are dominated by statistics and change only by 
10-15\% from model to model. 
We provide an approximate analytic expansion for the harmonic moments and their fluctuations
given in terms of the fixed-axes moments. For central collisions and in the absence of correlations 
it gives the simple formula 
$\Delta \varepsilon^\ast/\varepsilon^\ast \simeq \sqrt{4/\pi-1} = 0.52$.     
Similarly, we obtain expansions for the radial profiles of the higher harmonics. 
We investigate the relevance of the shape-fluctuation effects for jet quenching and find them
important only for very central events. 
Finally, we make some comments of relevance for hydrodynamics, the elliptic flow and its fluctuations. 
We argue how smooth hydrodynamics leads to the known result $v_4 \sim v_2^2$, and further to the prediction
$\Delta v_4/v_4 = 2 \Delta v_2/v_2$.
\end{abstract}

\pacs{25.75.-q, 25.75.Dw, 25.75.Ld}

\keywords{relativistic heavy-ion collisions, Glauber model, wounded nucleons, event-by-event fluctuations, elliptic flow}

\maketitle 

\section{Introduction\label{sec:intro}}

It was realized a few years ago in event-by-event hydrodynamic studies \cite{Aguiar:2000hw,Aguiar:2001ac} 
of relativistic heavy-ion collisions that  fluctuations of the initial shape
of the fireball formed in the early stage of the reaction lead to quantitatively important 
effects for azimuthal asymmetry \cite{Miller:2003kd,Bhalerao:2005mm,Andrade:2006yh,Voloshin:2006gz}. 
These effects, resulting from the shift of the center-of-mass and the rotation of the the quadrupole principal axis, 
can be seen in the analyses of the elliptic flow \cite{Alver:2006pn,Alver:2006wh,Sorensen:2006nw,Alver:2007rm}. 
The purpose of this paper is to investigate this phenomenon in detail
in the framework of various Glauber-like approaches describing the deposition of energy in the 
system in the early stages of the collision. Our study focuses on both the 
understanding of the statistical nature of the results, as well as on comparisons of various models.  
The main outcome presented in this paper is twofold: first, we provide the Fourier moments and radial 
profiles of the so-called {\em participant} type, {\em i.e.} obtained with an averaging procedure where 
in each event the system is translated to its center of mass and aligned with the major principal axis. 
Second, under reasonable approximations
we derive analytic expansions which explain the basic features of the Fourier moments and profiles.

\begin{figure*}[tb]
\begin{center}
\subfigure{\includegraphics[width=.31\textwidth]{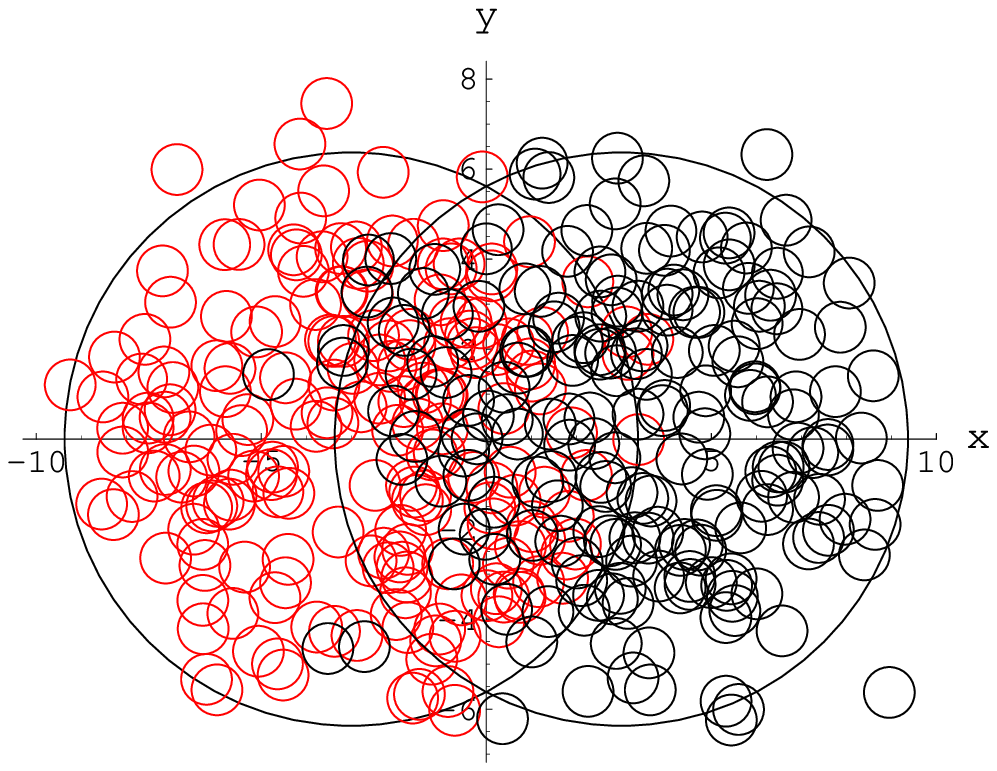}} 
\subfigure{\includegraphics[width=.31\textwidth]{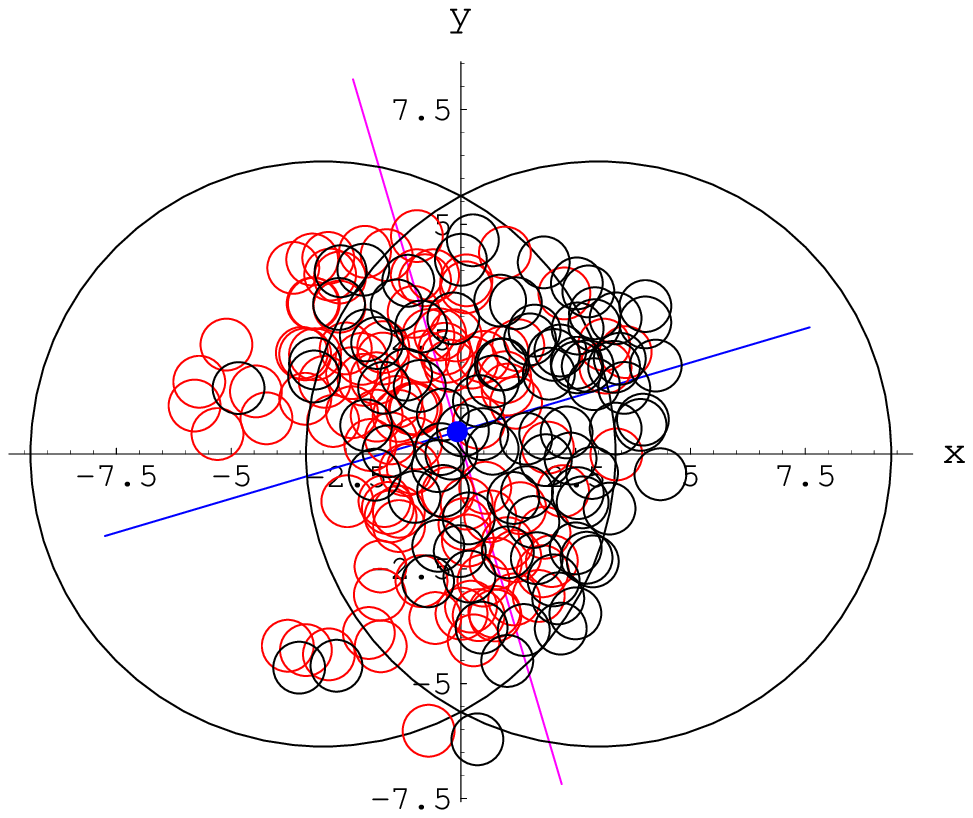}}
\subfigure{\includegraphics[width=.31\textwidth]{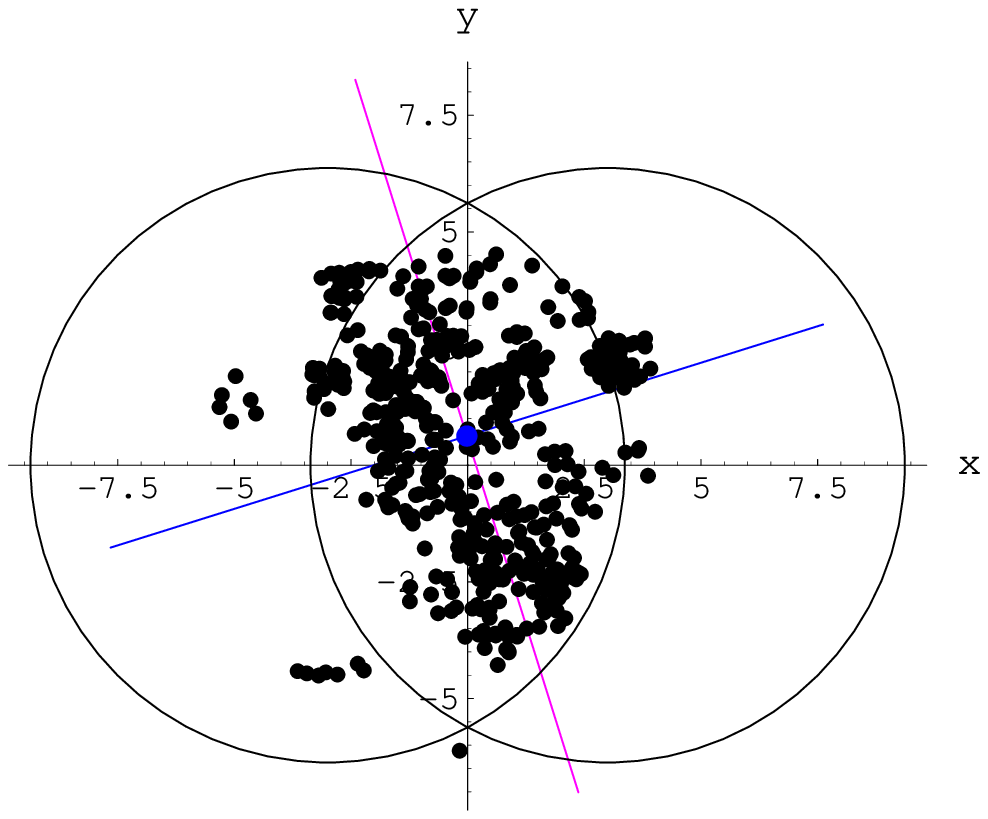}}
\end{center}
\caption{(Color online)
Snapshot of a typical gold-gold collision in the $x-y$ plane, $b=6$~fm. Red and black circles indicate 
nucleons from nuclei $A$ and $B$, respectively, plotted with the size (\ref{hard}). 
The left picture shows all nucleons, the middle - the wounded 
nucleons only, and the right - the centers of mass of pairs of nucleons undergoing binary collisions. The straight lines indicate 
the (twisted) principal axis of the quadrupole moment, the blue dots show the center of mass of the system, while the 
outer circles denote the Woods-Saxon radius of gold, $R=6.37$~fm. The units on the $x$ and $y$ axes are fm.
\label{fig:snap}}
\end{figure*}

The fact that the initial shape of the fireball fluctuates from event to event is certainly not 
surprising. 
Clearly, a finite number of sources consisting of wounded nucleons, binary collisions, {\em etc.},
which deposit the transverse energy in the system at mid-rapidity do not fill  
the available coordinate space uniformly due to statistical fluctuations. For instance, the center of mass 
of a system of uncorrelated particles fluctuates with a standard deviation proportional to $1/\sqrt{n}$. 
Similarly, the orientation of principal axis of the quadrupole and higher harmonic moments 
fluctuates from event to event. The statistical component of the harmonic moments 
assumes average values proportional $1/\sqrt{n}$, where $n$ 
is the number of sources. Since the number of sources is not so large, ranging from a few to a few hundred,
these fluctuations may easily reach a value of a few percent or higher, large for studies of azimuthal asymmetry where the 
investigated effects, such as the elliptic flow coefficient $v_2$, are of similar order.  
For the case of the wounded-nucleon model \cite{Bialas:1976ed} and for the binary collisions the 
situation is illustrated in Fig.~\ref{fig:snap}. The picture on the left shows all nucleons in both nuclei, 
the middle one the wounded nucleons, and the right one the binary collisions. 
We notice the mentioned effects for the distributions of the wounded nucleon and the binary collisions: 
the twist of the principal axes, denoted by the skewed lines, and the displacement of the center of mass, 
represented by a dot at the intersection of the principal axes. Statistical analyses may be carried out in the reference frame 
fixed by the reaction plane (we call it {\em fixed-axes}), or (in each event) in the frame defined by the shifted and twisted principal axes of 
the quadrupole moment (we call it {\em variable-axes}\footnote{We find this nomenclature more descriptive than the terms
{\em standard} and {\em participant} used in the literature.}).

The first part of the paper discusses the fixed-axes and variable-axes harmonic moments and radial profiles 
obtained numerically from the Glauber Monte Carlo studies in several models: 
the conventional wounded-nucleon model \cite{Bialas:1976ed}, 
a model admixing binary collisions to wounded nucleons \cite{Back:2001xy,Back:2004dy}, 
a model with {\em hot spots}, as well as 
the hot-spot model where the deposition of energy occurs with a given probability distribution (Sect.~\ref{sec:models}). 
The results are 
presented in Sect.~\ref{sec:fixed} and \ref{sec:var}.
The main result here is that the {fixed-axes} quadrupole moments, $\varepsilon$, and their scaled 
standard deviation,  $\Delta \varepsilon/\varepsilon$, vary significantly from model to model.
The same holds to a lesser extent for the {variable-axes} moments, $\varepsilon^\ast$.
On the other hand, the dependence of the scaled standard deviation $\Delta \varepsilon^\ast/\varepsilon^\ast$ 
on the chosen Glauber-like model is weak, at most at the level of 10-15\% for intermediate impact parameters. 
For all considered models the values range from about 0.5 for central collisions to about 0.3-0.4 for peripheral collisions.
We examine the dependence on the mass number, 
providing results for the gold-gold and copper-copper collisions. 
We also investigate the effects of the assumed weighting power of the transverse radius in the definition of the 
harmonic moments, finding that the choice is not important for studies of fluctuations.

In Sect.~\ref{sec:jets} we examine the role of the center-of-mass and quadrupole-axes fluctuations on jet quenching. 
Except for very central collisions, the effect of 
the increased eccentricity of the 
opaque medium is canceled by the shift of its position and axes rotation,
leading to almost no change in the 
azimuthal asymmetry of the jets leaving the interaction region.

In Sect.~\ref{sec:moments} we argue that the variable-axes quantities are dominated by sheer statistics and 
certain properties of variable-axes distributions can be explained in an elementary way through the 
use of the central limit theorem. In particular, 
in the absence of correlations between the location of sources and for central collisions we get the result of an
appealing simplicity, namely 
\mbox{$\Delta \varepsilon^\ast/ \varepsilon^\ast(b=0)=\sqrt{4/\pi-1}\simeq 0.52$}, independent of the number of sources in the 
assumed model, the mass number of the colliding nuclei, or the collision energy. 
This result is fulfilled to a very good accuracy in actual numerical studies, where some correlations are present.
For non-central collisions appropriate expansions are provided. We also analyze the variable-axes profiles 
in this way. The effects of correlations between the location of sources are discussed in Appendix~\ref{app:case2}.   

In Sect.~\ref{sec:ee} we propose another method of encoding the 
information on the initial state, where each harmonic (including the odd ones) is evaluated 
in its own eigen-axes. The method can be used as a base for a smoothing 
procedure in preparation of the initial conditions for event-by-event hydrodynamic
studies.

In Sect.~\ref{sec:flow} we make several comments referring to the collective 
flow. We note that the statistical analysis of the variable-axes 
parameters $\varepsilon^\ast$ 
carries over to the analysis of the 
variable-axes elliptic-flow coefficient, $v_2^\ast$. 
For central collisions (in the absence of 
correlations) we find \mbox{$\Delta v_2^\ast/v_2^\ast(b=0)=\sqrt{4/\pi-1} \simeq 0.52$}, independently of multiplicity, 
mass number, or the collision energy. This value is in the 
ball park of the recent experimental data \cite{Sorensen:2006nw,Alver:2007rm}.
Moreover, under the assumption of smoothness that most likely holds in hydrodynamics, which allows for 
perturbation theory around the azimuthally symmetric solution, one obtains the relation
$v_4^\ast \sim v_2^{\ast 2}$ for the octupole flow coefficient. Consequently, for the event-by-event fluctuations we find
the prediction \mbox{$\Delta v_4^\ast/v_4^\ast=2\Delta v_2^\ast/v_2^\ast$}.

Appendices contain some more technical material, including the derivations of the statistical 
formulas. A simple one-dimensional toy model illustrating the 
essence of the statistical intricacies is given in Appendix~\ref{sec:toy}.

\section{Notation\label{sec:notations}}

In our study we use the standard Woods-Saxon nuclear density profile for the nucleus of mass number $A$,
\begin{eqnarray}
n(r)=\frac{c}{1+\exp(\frac{r-R}{a})}, \label{ws}
\end{eqnarray}
where the constant $c$, given in Appendix \ref{app:lowhighr},
is such that the normalization 
$\int 4\pi r^2 dr \,n(r)=A$ is fulfilled. 
For the considered gold and copper nuclei the parameters are
\begin{eqnarray}
&&R=6.37~{\rm fm},\;\; a=0.54~{\rm fm}, \;\;\; ({}^{197}{\rm Au}),  \label{gold} \\ 
&&R=4.14~{\rm fm},\;\; a=0.57~{\rm fm}, \;\;\; ({}^{62}{\rm Cu}). \label{copper} 
\end{eqnarray}
A popular way to simulate the short-range repulsion in Glauber-like calculations\footnote{We note that this 
repulsion increases slightly the size of the nucleus, but the effect is negligible, see Appendix~\ref{app:monte}.} 
is to enforce that the centers of nucleons in each 
nucleus cannot be closer to each other than the expulsion distance of $d=0.4$~fm. 
This feature is simple to implement in 
Monte Carlo generators. Some details are provided in Appendix~\ref{app:monte}.

We use the following standard convention for the axes of the reference frame: the $z$-axis is along the beam, the $x$-axis 
lies in the reaction plane,
and the $y$-axis is perpendicular to the reaction plane. The azimuthal angle $\phi \in [-\pi,\pi]$
is measured relative to the $y$-axis, thus $y=\rho \cos \phi$,  $x=\rho \sin \phi$, where $\rho$ is the transverse radius. 

We refer to the analysis in the fixed reference frame of the reaction plane as {\bf fixed-axes} (sometimes called {\em standard} in 
the literature), and to the analysis where the particles in each event are translated to the center-of-mass frame and aligned with the 
major principal axis of the quadrupole moment as {\bf variable-axes} (also called {\em participant}).  

\section{Models\label{sec:models}}

We describe briefly the models studied in this paper.
The standard implementation of the 
wounded nucleon model at RHIC energies assumes that the inelastic cross 
section of the nucleon is 
\begin{eqnarray}
\sigma_{w}=42~{\rm mb}. \label{swound}
\end{eqnarray}
The nucleon from one nucleus gets wounded when it passes closer to a nucleon 
from 
the other nucleus than the hard-sphere radius
\begin{eqnarray}
r_0=\frac{1}{2}\sqrt{\sigma_{w}/\pi}. \label{hard}
\end{eqnarray} 
Then the weight $w=1/2$ is attributed to the point in the 
 transverse plane at the position of the wounded nucleon.
The weight can be thought of as a measure 
proportional to the amount of the deposition of the transverse energy, 
which then 
is carried away by the produced particles.
For studies of fluctuations only the relative weights are important, and the 
overall 
normalization of the total weight can be chosen arbitrarily. In what follows 
we renormalize the distributions in all models to the number of the 
wounded nucleons, $N_w$.

For binary collisions the weight $w=1$ is attributed to each collision point, 
which is taken as the mean of the coordinates of the 
two colliding nuclei.

A successful description of multiplicities at RHIC has been achieved with a {\em mixed} model, 
amending the wounded nucleon model  \cite{Bialas:1976ed}
with some binary collisions \cite{Back:2001xy,Back:2004dy}. 
In this case a wounded nucleon obtains the weight $w=(1-\alpha)/2$, and a binary collision 
the weight $w=\alpha$. The total weight averaged over events 
is then $(1-\alpha)N_{\rm w}/2+\alpha N_{\rm bin}$. The fits to particle 
multiplicities of Ref.~\cite{Back:2004dy} give $\alpha = 0.145$ for
collisions at $\sqrt{s_{NN}}=200$~GeV, and $\alpha = 0.12$ for $\sqrt{s_{NN}}=19.6$~GeV.

Next, we consider a model with {\em hot spots} in the spirit of Ref.~\cite{Gyulassy:1996br}, assuming that the 
cross section for a semi-hard binary collisions producing a hot-spot 
is small, $\sigma_{\rm hot-spot} = 0.5$~mb, but when such a rare collision occurs it produces 
on the average a large amount of transverse energy equal to $\alpha\sigma_{\rm w}/\sigma_{\rm hot-spot}$. 

Each source from the previous models (wounded nucleon, mixed, or hot-spot) may 
deposit the transverse energy with a certain probability distribution. To incorporate this effect, we
superimpose the $\Gamma$ distribution over the distribution of sources, multiplying the weights of the considered model with the 
randomly distributed number from the gamma distribution 
\begin{eqnarray}
g(w,\kappa)=\frac{w^{\kappa-1}\kappa^\kappa \exp(-\kappa w)}{\Gamma(\kappa)}.
\end{eqnarray} 
This distribution gives the average value equal to $\bar w = 1$ and the variance  ${\rm var}(w)= 1/\kappa$. This is at no loss 
of generality, since, as already mentioned, the individual weights used for carrying the statistical averages 
can be normalized arbitrarily. 
In this paper we do this superposition on the hot-spot model, where the considered effects are largest.
Thus, we take the weights $(1-\alpha)g(w,\kappa)/2$ for the wounded nucleons and  $\alpha g(w,\kappa)\sigma_{\rm w}/\sigma_{\rm hot-spot}$ 
for the binary collisions. We take $\kappa=0.5$, which gives ${\rm var}(w)=5$.
We label this model {\em hot-spot+$\Gamma$}.

The four considered models, {\bf wounded nucleon}, {\bf mixed}, {\bf hot spot}, and 
{\bf hot-spot+$\Gamma$}, differ by the number of sources and the amount of the built-in fluctuations. 
For instance, the superposition of the $\Gamma$ distribution with low values of $k$ increases the variance. 
This increase is also generated with hot spots, which 
effectively reduce the number of sources. All these effects will be studied in detail below.
    
\section{Fixed-axes harmonic moments\label{sec:fixed}}

When the reaction plane is determined for each event (which of course can never be achieved 
exactly in the experiment, see {\em e.g.} Refs.~\cite{Poskanzer:1998yz,Ollitrault:1992bk}), one can then 
choose the reference frame fixed by the reaction plane. 
The two-dimensional (boost-invariant) profile of the density of sources, $f(\rho,\phi)$, 
is obtained by averaging over the events belonging to a particular centrality or impact parameter class.  
The symmetry $f(\rho,\phi)=f(\rho,\pi-\phi)$ excludes odd components in the Fourier decomposition, while 
for equal colliding nuclei the symmetry $f(\rho,\phi)=f(\rho,-\phi)$ eliminates the $\sin(l \phi)$ functions.
Thus,
\begin{eqnarray}
f(\rho,\phi)=f_0(\rho) + 2 f_2(\rho) \cos(2\phi)+ 2 f_4(\rho) \cos(4\phi)+ \dots, \nonumber \\ \label{f}
\end{eqnarray}
where $\rho$ is measured from the center of the geometric intersection of the two nuclei. 
The harmonic moments obtained form (\ref{f}), which we call {\em fixed-axes}, are also called ``standard'' 
in the literature.

We first have a look at the Fourier profiles $f_l(\rho)$, where $l=0,2,4,\dots$.
In Appendix \ref{app:lowhighr} we show that at low values of $\rho$
\begin{eqnarray}
f_l(\rho) \sim \rho^l, \;\;\;\;(\rho \ll b), \label{lowr}
\end{eqnarray} 
while at high values of $\rho$
\begin{eqnarray}
f_l(\rho) \sim \exp(-2\rho/a) \rho^2 \left ( \frac{b}{\rho} \right )^l, \;\;\;(\rho \gg b). \label{highr}
\end{eqnarray} 
Such a behavior is typical of Fourier expansions of smooth functions. 

\begin{figure}
\begin{center}
\subfigure{\includegraphics[width=.5\textwidth]{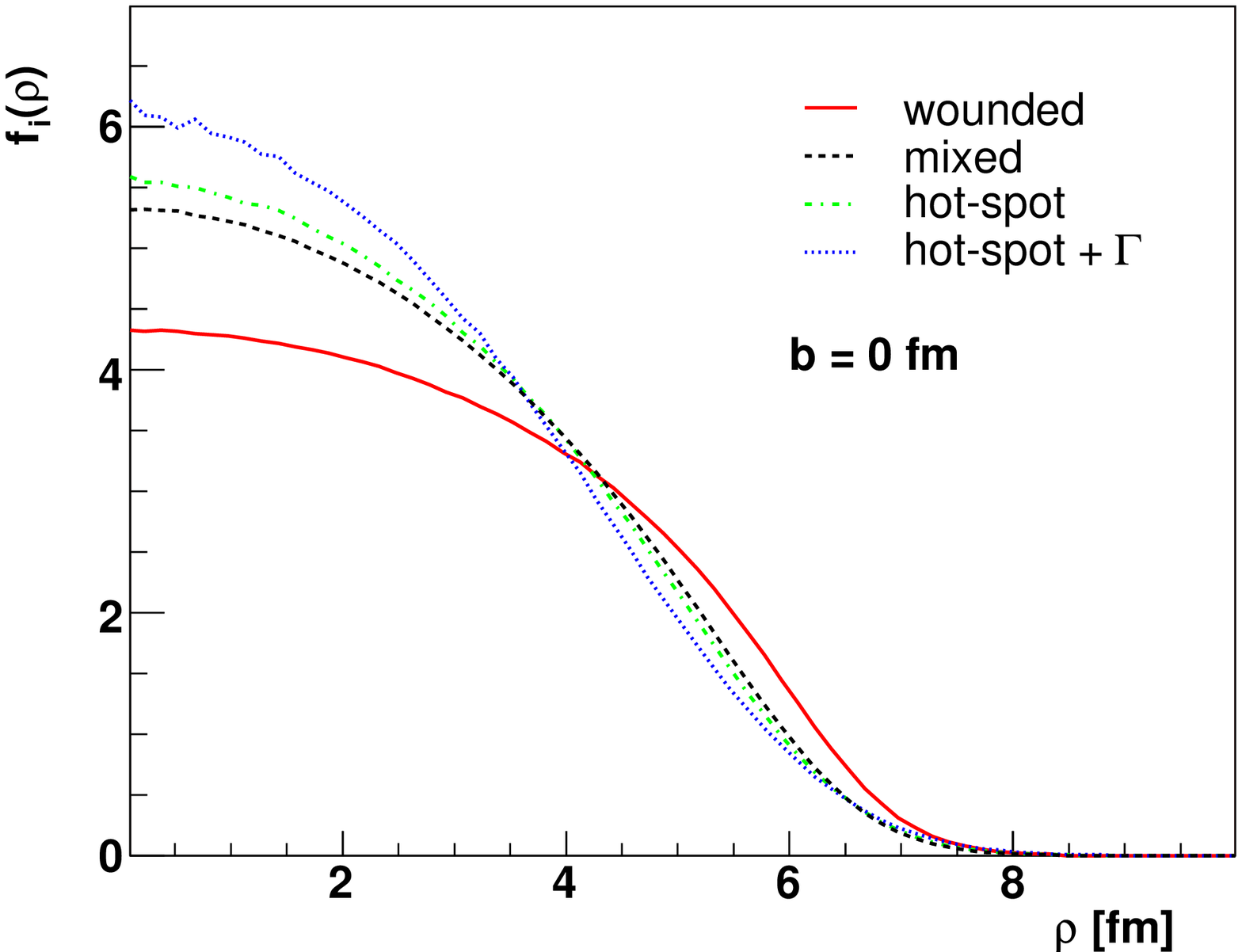}}\\ 
\subfigure{\includegraphics[width=.5\textwidth]{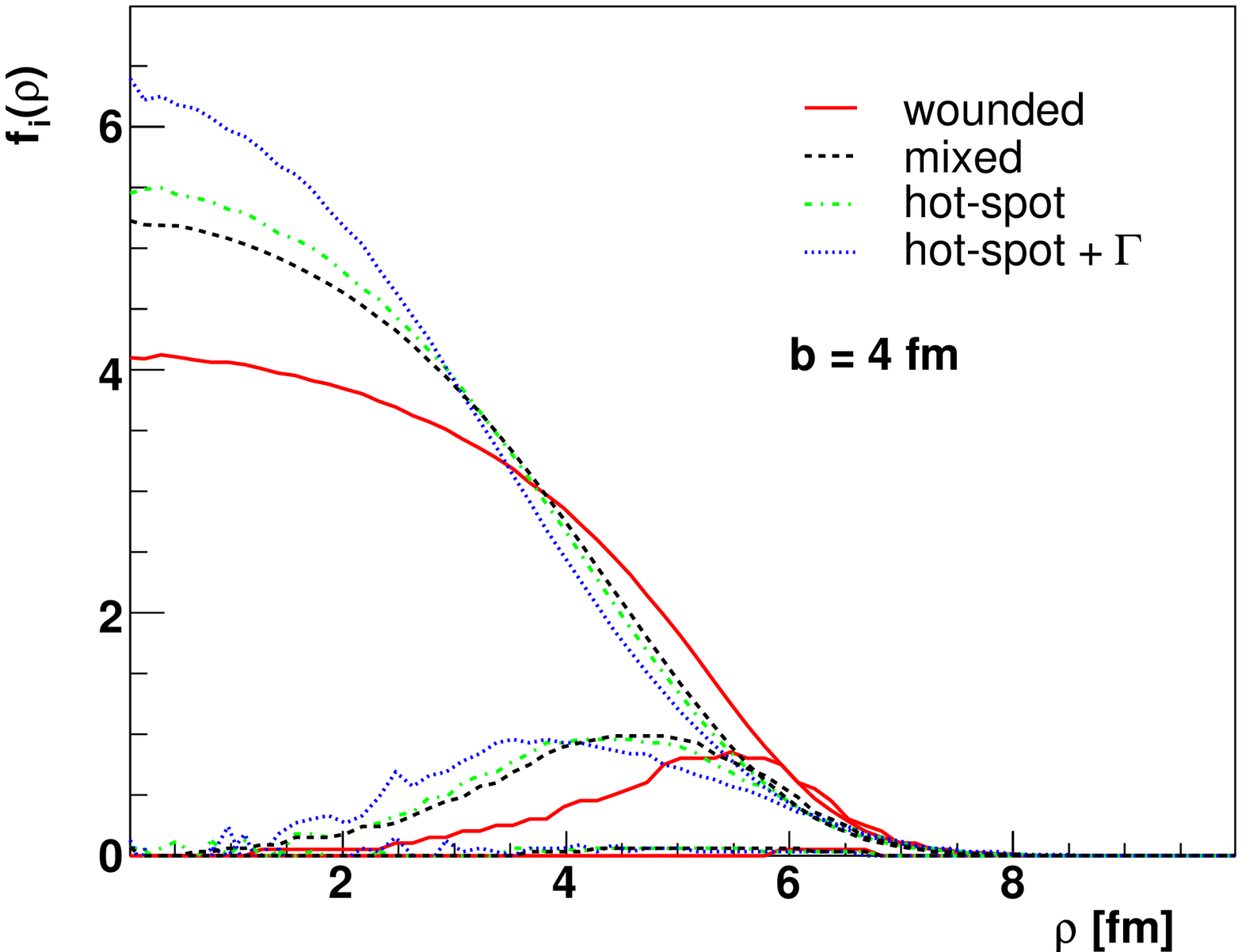}}\\
\subfigure{\includegraphics[width=.5\textwidth]{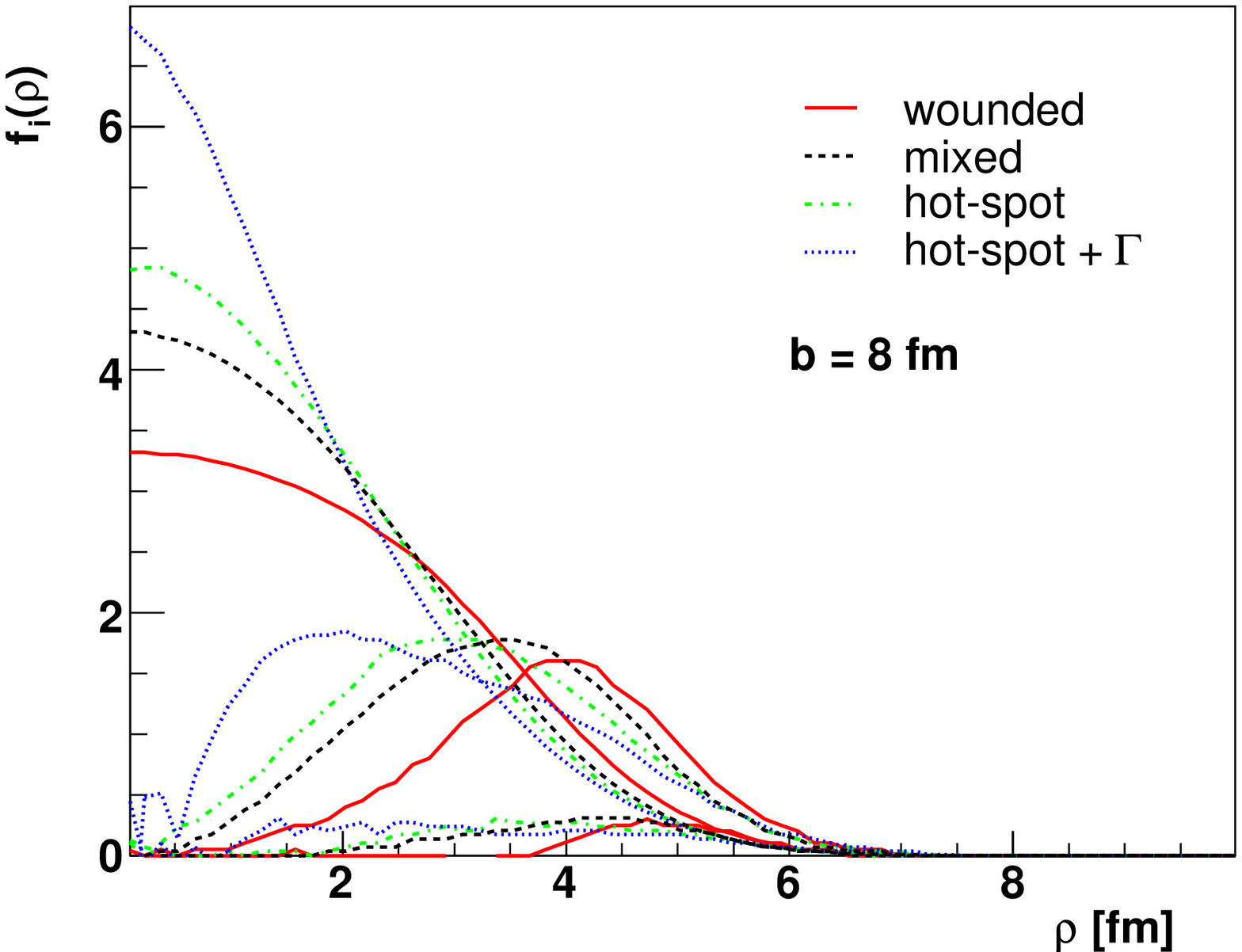}}\\
\end{center}
\caption{(Color online) The {\em fixed-axes} profiles $f_i(\rho)$ for gold-gold collisions in the analyzed models at several values of the 
impact parameter: $b=0, 4, 8$~fm. The legend explains the assignment of line-types to the
models. For $b=4$ and $8$~fm for each model there are three curves, corresponding to 
$i=0$, $2$, and $4$ from top to bottom. The $i=4$ component is tiny. \label{fig:profiles}}
\end{figure}

We present the profiles $f_l(\rho)$ obtained with Monte Carlo simulations for the considered models in Fig.~\ref{fig:profiles}.
The distributions in all models are normalized to the number of wounded nucleons, {\em i.e.}
\begin{eqnarray}
\int \rho\,d\rho\,d\phi f(\rho,\phi)= \int 2\pi \rho d \rho f_0(\rho)=N_w. \label{normnw} 
\end{eqnarray} 
We note that the profiles are substantially different from model to model. The monopole profile $f_0$ is broadest in 
the wounded nucleon model, then passing through the mixed model and the hot-spot model we arrive at the  hot-spot+$\Gamma$ model, 
which is most sharply peaked at the origin. Correspondingly, the quadrupole profiles $f_2(\rho)$ are concentrated 
further or closer to the origin. We note that the amplitude of subsequent harmonics decreases fast, such that taking $l$ up to 4 
is sufficient for any practical matter.  

In order to have some convenient quantitative measures of the profiles of Eq.~(\ref{f})
one introduces their radial moments
\begin{eqnarray}
\varepsilon_{k,l}=
\frac{\int 2\pi \rho f_l(\rho) \rho^k d\rho }{\int 2\pi \rho f_0(\rho) \rho^k
d\rho }=\frac{I_{k,l}}{I_{k,0}}.
\label{epsilon}
\end{eqnarray}
where we have introduced the moments
\begin{eqnarray}
I_{k,l}=\frac{1}{N_w} \int_0^\infty 2\pi \rho d\rho f_l(\rho) \rho^k
\end{eqnarray}
for a future reference.
The choice of the weighting power $k$ is arbitrary, with the typical choice $k=2$. 
Higher values of $k$ make the measure more sensitive to the outer region of the system. 
We note that in the popular notation 
\begin{eqnarray}
\varepsilon_{\rm std}=\varepsilon_{2,2}\equiv \varepsilon.
\end{eqnarray} 

\begin{figure}
\includegraphics[width=.5\textwidth]{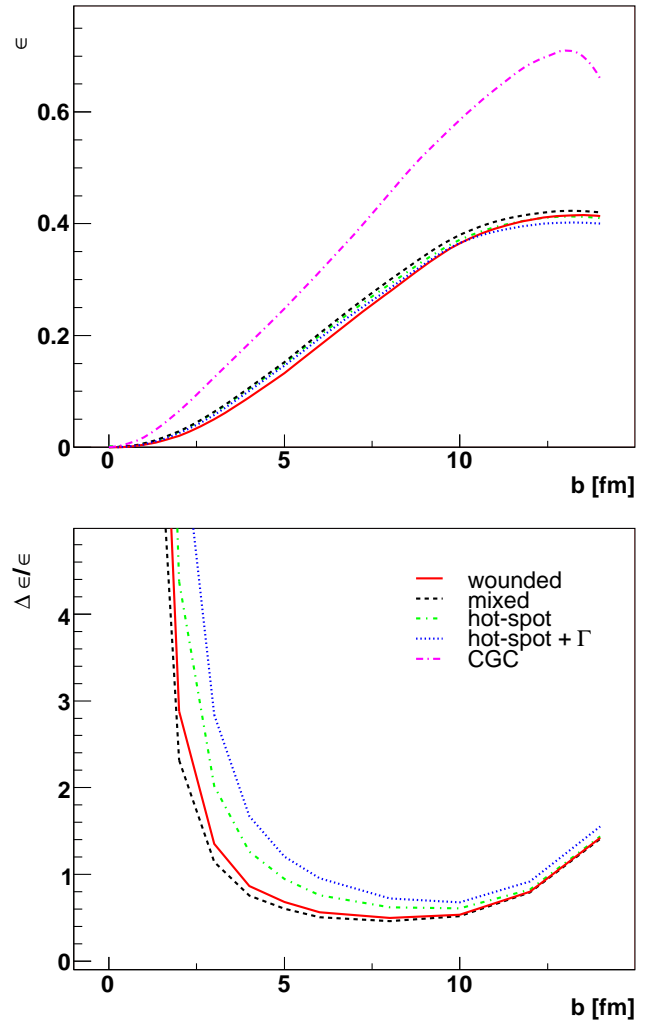}
\caption{(Color online) The harmonic moment $\varepsilon\equiv\varepsilon_{2,2}$ and its scaled standard deviation 
for the analyzed models
plotted as functions of the impact parameter. The result for the color-glass 
condensate comes from Ref.~\cite{Hirano:2005xf}. Gold-gold collisions. \label{fig:epsilons}}
\end{figure}

We observe that in all the considered Glauber models $\varepsilon$ is practically independent of the model (top panel of 
Fig.~\ref{fig:epsilons}). Tiny differences come from different distributions of the wounded nucleons and 
binary collisions. 
On the other hand, the scaled standard deviation, shown at the lower
panel of Fig.~\ref{fig:epsilons}, displays a strong dependence on the model at low values of $b$,
with the hot-spot+$\Gamma$ model yielding about twice as much as the mixed model. 
We also notice a strong dependence on $b$. At $b=0$ the curves diverge, which is an artefact of dividing by the vanishing 
value of $\varepsilon$.
The fluctuations are larger in models effectively having the lower number of sources, which is obvious 
from the statistical point of view.

As already noted in Refs.~\cite{Hirano:2005xf,Drescher:2006ca}, 
the value of $\varepsilon$ obtained with the color glass condensate (CGC)  
is substantially higher than in all Glauber-like models analyzed in this paper. For comparison, the CGC result is 
shown as the upper curve in the top panel of Fig.~\ref{fig:epsilons}. 
After the e-print version of this paper has been posted, a calculation of  
fluctuations of $\varepsilon^\ast$ in the CGC framework has appeared \cite{DNara}.
The results are overlayed in in the bottom part Fig.~\ref{fig:epsilonsa}. We note that 
at intermediate values of $b$ the CGC values 
of $\Delta \varepsilon^\ast/\varepsilon^\ast$ are significantly lower than in the 
considered Glauber models. 


\begin{figure}
\includegraphics[width=.5\textwidth]{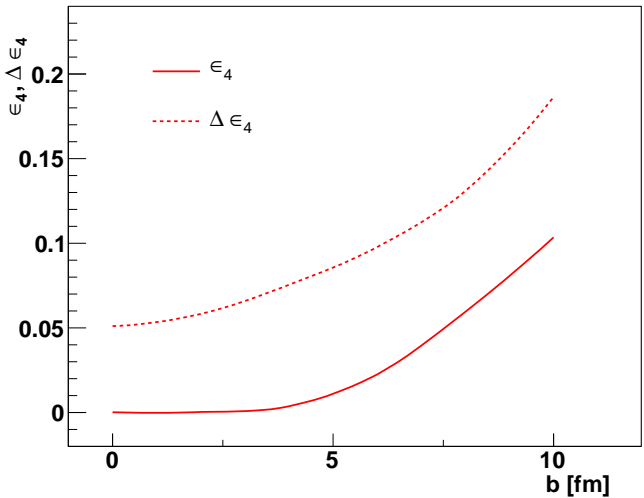}
\caption{(Color online) The fixed-axes octupole moment, $\varepsilon_{4}\equiv\varepsilon_{2,4}$, and 
its standard deviation, plotted as functions of the impact parameter. Wounded nucleon model, gold-gold
collisions. \label{fig:eps4_s}}
\end{figure}

In Fig.~\ref{fig:eps4_s} we show the results for the octupole moment, $\varepsilon_{4,2}$, and 
its standard deviation, obtained in the wounded-nucleon model. We note a very flat shape of $\varepsilon_{4,2} \sim b^4$ at low $b$.
This behavior is a direct consequence of integrating Eqs.~(\ref{lowr},\ref{highr}).
The standard deviation grows with increasing impact parameter. In other models considered in this paper the results are
qualitatively similar.

\section{Variable-axes harmonic moments\label{sec:var}}

As is well known, 
the reaction plane cannot be determined precisely in an experiment, 
or not at all, which originated multiple methods of analyzing
azimuthal asymmetry in heavy-ion collisions. 
As has recently been realized \cite{Aguiar:2000hw,Aguiar:2001ac,Miller:2003kd,Bhalerao:2005mm,Andrade:2006yh,Voloshin:2006gz}, 
the purely statistical fluctuations caused 
by the finite number of particles lead to sizeable effects of the {\em variable geometry} in the initial stage of the collision. 
The effect can be seen qualitatively in Fig.~\ref{fig:snap}, where we notice a highly irregular shape of 
the distributions of  both the wounded nucleons
and the binary-collisions distribution.
The mere presence of the fluctuation of the initial condition is obvious. 
What is somewhat surprising, however, is its size, leading to noticeable
effects in the analysis of azimuthal asymmetry even at large numbers of participating nucleons. 

\begin{figure}
\includegraphics[width=.5\textwidth]{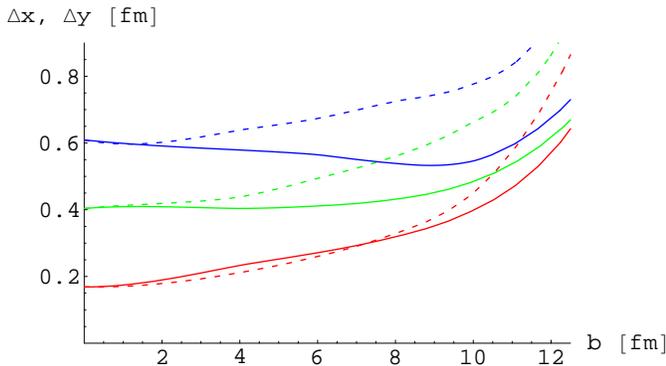}
\caption{(Color online) The root mean square shifts of the center of mass in the 
in-plane direction, $\Delta x$ (solid lines), and in the out-of-plane 
direction, $\Delta y$ (dashed lines). The lower lines are for the wounded-nucleon, the middle 
for the hot-spot, and the upper for the hot-spot+$\Gamma$ model. \label{fig:shifts}}
\end{figure}

\begin{figure}
\begin{center}
\subfigure{\includegraphics[width=.5\textwidth]{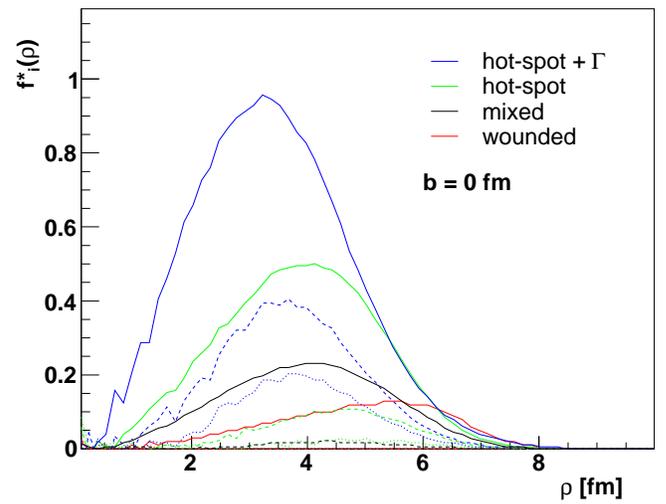}}\\ 
\subfigure{\includegraphics[width=.5\textwidth]{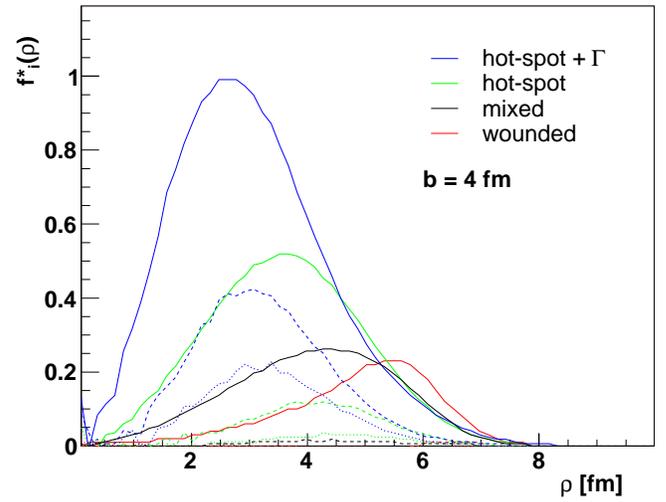}}\\
\subfigure{\includegraphics[width=.5\textwidth]{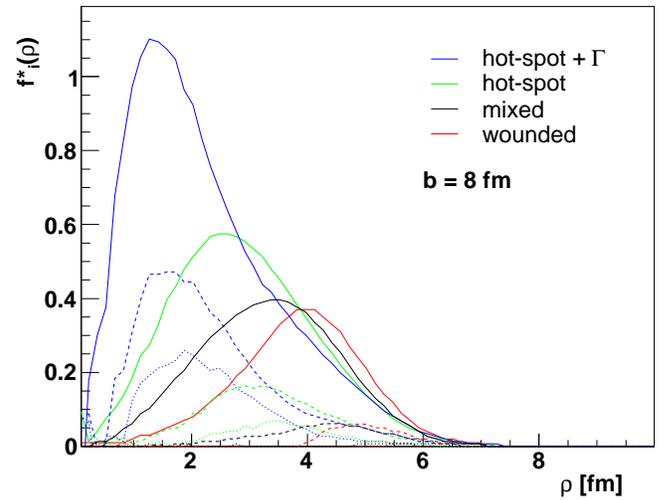}}\\
\end{center}
\caption{(Color online) The {\em variable-axes} profiles $f^\ast_i(\rho)$ for the analyzed models at several values of the 
impact parameter. The solid, dashed, and dotted lines correspond to $i=2$, $4$, and $6$, respectively. 
The four models: hot-spot+$\Gamma$, hot-spot, mixed, and the wounded nucleon model 
yield curves arranged from the top to bottom, respectively. 
Gold-gold collisions. \label{fig:profilesa}}
\end{figure}

The center of mass of the distribution of the sources  is not located at the geometrical center 
of the overlap of the two-colliding nuclei. In each event it is shifted in the $x$ and $y$ directions by 
$\Delta x$ and $\Delta y$, defined as 
\begin{eqnarray}
\Delta x = \frac{\sum_i x_i w_i}{\sum_i w_i}, \;\; \Delta y = \frac{\sum_i y_i w_i}{\sum_i w_i},.
\end{eqnarray}
where $i$ labels the sources and $w_i$ denotes the weights.
For an uncorrelated distribution of a large number of sources $n$ one has 
\begin{equation} 
(\Delta x)^2=\frac{1}{{n}} R_x^2 \langle w^2 \rangle, \;\; (\Delta y)^2=\frac{1}{{n}} R_y^2 \langle w^2 \rangle, \label{xy}
\end{equation}
where
\begin{eqnarray}
&&R_x^2 = \frac{1}{N_w} \int f(\rho,\phi) \rho^3 \sin^2\phi d\rho d\phi, \nonumber \\ 
&&R_y^2 = \frac{1}{N_w} \int f(\rho,\phi) \rho^3 \cos^2\phi d\rho d\phi,
\end{eqnarray}
are the mean squared radii of the geometric fixed-axes distribution, and 
$\langle w^2 \rangle = \int dw\, w^2 g(w,\kappa)$.
Since our system exhibits some correlation between the location of sources, the formula (\ref{xy})
is not realized exactly, but hold qualitatively.
In Fig \ref{fig:shifts} we show $\Delta x$ and $\Delta y$ 
as functions of the 
impact parameter for the wounded-nucleon (lower curves), the hot-spot (middle curves) and the hot-spot+$\Gamma$ (top curves) models. 
The shift is  more important for peripheral collisions, 
where $n$ is lower, even though the source size itself 
decreases. 

The shift of the center of mass is physically relevant in the jet analysis, because it moves apart the formed
fireball from the jet production points (Sect.~\ref{sec:jets}).

In each event one can compute the principal axes of the ellipse of inertia. This corresponds to the 
choice $k=2$ as the weighting power. The angle between the major half-axis of the ellipse and the $y$ 
axis is given by
\begin{eqnarray}
\tan (2\phi^\ast) = 2 \frac{\langle xy \rangle - \langle x \rangle \langle y \rangle}{{\rm var}(y)-{\rm var}(x)}, \label{tanast}
\end{eqnarray}  
where the brackets denote the averaging over the particles within the given event. 
Importantly, in the variable-axes calculations all coordinates in the given event are always  
shifted by $(\Delta x,\Delta y)$ such that $\langle x \rangle=\langle y \rangle=0$.

The angle $\phi^\ast$ fluctuates sizably from event to event.
In our notation, the superscript $\ast$ indicates quantities averaged in such a way, that first in each event the rotation angle $\phi^\ast$ is determined
according to Eq.~(\ref{tanast}), then the
rotation is performed to 
the current principal-axis system, and finally the summation is done. As a result, 
\begin{eqnarray}
f^\ast(\rho,\phi)&=&f^\ast_0(\rho) + 2 f^\ast_2(\rho) \cos(2\phi-2\phi^\ast) \nonumber \\ 
&+& 2 f^\ast_4(\rho) \cos(4\phi-4\phi^\ast)+ \dots \label{fr}
\end{eqnarray}
We call the above profiles the {\em variable-axes} profiles.
Obviously, $f^\ast_0(\rho)=f_0(\rho)$ and  $f^\ast_l(\rho) \ge f_l(\rho)$ for each $\rho$.

\begin{figure}
\includegraphics[width=.5\textwidth]{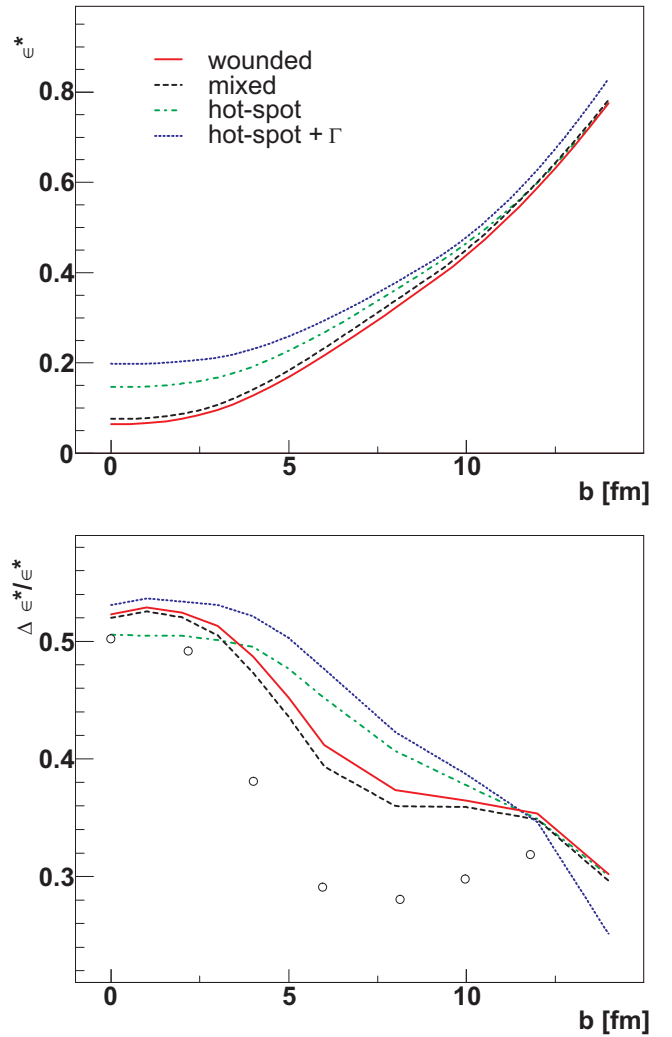}
\caption{(Color online) The harmonic moment $\varepsilon^\ast \equiv \varepsilon^\ast_{2,2}$ and its scaled standard deviation 
for the analyzed models
plotted as functions of the impact parameter. Gold-gold collisions. The open circles in the bottom figure show the calculation 
in the color-glass-condensate framework taken from Ref.~\cite{DNara}.
 \label{fig:epsilonsa}}
\end{figure}

In analogy to Eq.~(\ref{epsilon}) we introduce the variable-axes moments
\begin{eqnarray}
\varepsilon^\ast_{k,l}=\frac{\int 2\pi \rho f^\ast_l(\rho) \rho^k}{\int 2\pi \rho f^\ast_0(\rho) \rho^k}, \label{epsilona}
\end{eqnarray}
In the common notation for the variable-axes or participant deformation parameter one has
\begin{eqnarray}
\varepsilon_{\rm part}=\varepsilon^\ast_{2,2} \equiv \varepsilon^\ast. \label{epart} 
\end{eqnarray}
The profiles and moments for higher harmonics are suppressed, similarly to the fixed-axes case. 
This is clear, as the higher harmonics are evaluated relative to the axes determined by maximizing 
the quadrupole moment. As a result, only a few moments are needed to effectively parameterize the profile.

Figure \ref{fig:profilesa} shows the variable-axes profiles for $l=2,4,6$ (the $l=0$ profiles are equal to the 
fixed-axes case) for all considered models. The line-types distinguish the Fourier label $l$, while the 
hot-spot+$\Gamma$, hot-spot, mixed, and the wounded nucleon models yield curves for each $l$ arranged from the top 
to bottom, respectively.
Comparing Figs.~\ref{fig:profiles} and \ref{fig:profilesa} we note sizeable departures 
of the variable-axes profiles $f_l^\ast$ from the fixed-axes profiles $f_l$ ($l=2,4,\dots$), in particular at 
small values of the impact parameter.
For the central collisions ($b=0$) the variable-axes 
profiles are non-zero solely due to fluctuations, as will be discussed in Sect.~\ref{sec:moments}.  

In Fig.~\ref{fig:epsilonsa} we show the quadrupole moment $\varepsilon^\ast$ and its scaled standard deviation.
We observe a strong model dependence of $\varepsilon^\ast$ at low values of $b$, with models having effectively lower number 
of sources yielding higher values. At $b=0$ the hot-spot+$\Gamma$ model yields three times more than the 
wounded nucleon model. For all models the scaled standard deviation is close to the value
0.5 for central collisions and drops to about 0.3 at $b=14$~fm. 
We argue in Sect.~\ref{sec:moments} why the central value is always close to 0.5, independently of the effective 
number of sources.
At intermediate values of $b$ the relative difference in $\Delta \varepsilon^\ast/\varepsilon^\ast$ between various
considered models is at the level of 10-15\%, which is not a very strong effect.

We have examined numerically the role of the weighting power of the radius, $k$, entering the definitions (\ref{epsilon},\ref{epsilona}).
This is important for the method, since as pointed out earlier, the value of $k$ is arbitrary.
The result is that both $\varepsilon_{k,2}$ and $\varepsilon^\ast_{k,2}$ increase substantially with $k$, however 
the scaled standard deviation remains quite stable, in particular at low impact parameters. 
The results are collected in Table~\ref{tabb}. We vary $k$ between 0 and 6, which is a wide range. 
At $b=0$  the scaled standard deviation remains practically 
constant, while at $b=8$ it varies by 10\% for the fixed-axes case and 25\% for the variable-axes case.
Due to this rather weak dependence, 
the particular choice of the weighting power $k$ 
is not essential in studies of this quantity in event-by-event fluctuations.
However, the values of $\varepsilon$ and  $\varepsilon^\ast$ itself are sensitive to the choice of $k$.
As already mentioned, the higher values of $k$ increase the sensitivity to the profiles at higher 
values of the transverse radius $\rho$.

\begin{table}
\caption{Dependence of the quadrupole asymmetry parameters 
on the weighting power $k$ from definitions (\ref{epsilon},\ref{epsilona}) \label{tabb}}
\begin{tabular}{|l|cccc|}
\hline
$k$ & 0 & 2 & 4 & 6\\
\hline
\multicolumn{5}{|c|}{$b=0$}\\
\hline
$\varepsilon_{k,2}^\ast$                         & 0.047 & 0.064 & 0.089 & 0.121 \\ 
$\Delta \varepsilon_{k,2}^\ast/\varepsilon_{k,2}^\ast$ & 0.53 & 0.52 & 0.52 & 0.52 \\
\hline
\multicolumn{5}{|c|}{$b=8$~fm}\\
\hline
$\varepsilon_{k,2}$                         & 0.147 & 0.278 & 0.388 & 0.466 \\ 
$\Delta \varepsilon_{k,2}/\varepsilon_{k,2}$      & 0.44 & 0.48 & 0.48 & 0.49 \\
\hline
$\varepsilon_{k,2}^\ast$                         & 0.176 & 0.319 & 0.452 & 0.555 \\ 
$\Delta \varepsilon_{k,2}^\ast/\varepsilon_{k,2}^\ast$ & 0.44 & 0.38 & 0.34 & 0.31 \\
\hline
\end{tabular}
\end{table}

\begin{figure}
\includegraphics[width=.5\textwidth]{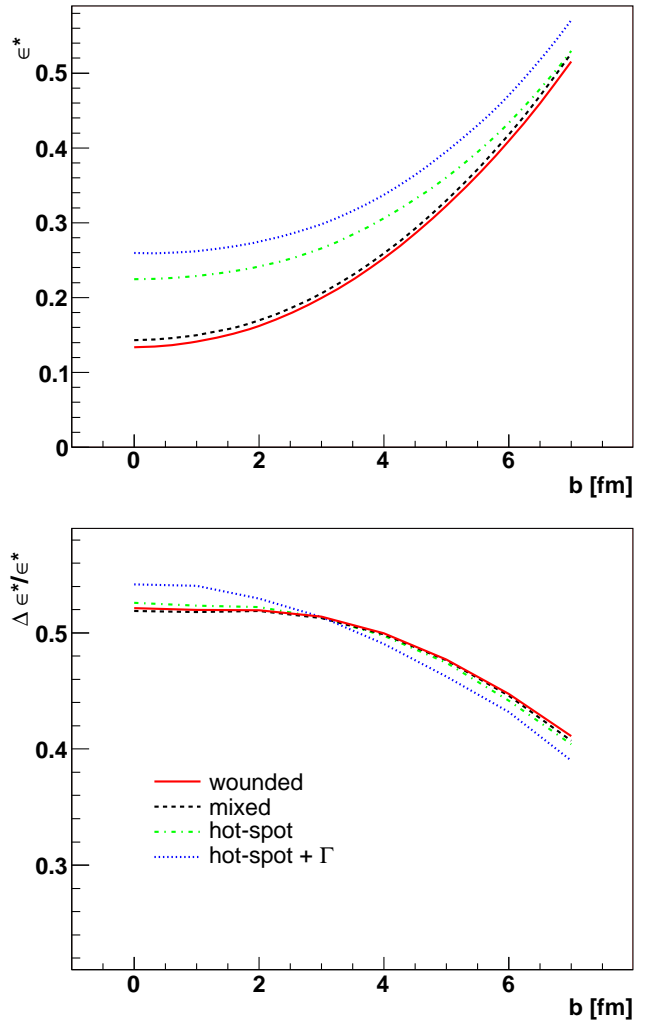}
\caption{(Color online) Same as Fig.~\ref{fig:epsilonsa} for the copper-copper collisions. \label{fig:copper}}
\end{figure}

In Fig.~\ref{fig:copper} we present the variable-axes harmonic moments and their scaled standard deviation 
for the copper-copper collisions. Due to the  much lower number of sources compared to the gold-gold case of Fig.~\ref{fig:epsilonsa},
we note higher values of $\varepsilon^\ast$ at low $b$. On the other hand, the scaled standard deviation is remarkably similar
to the gold-gold case, especially at low $b$, as should be according to the arguments of Sect.~\ref{sec:moments}. 
Certainly, the dependence on the mass number is a sensitive probe of the whole approach.

In Sect.~\ref{sec:moments} we will show that many of the qualitative and quantitative features of the 
Fourier distributions as well as their moments have simple explanations on purely statistical 
grounds.

\section{Jet quenching\label{sec:jets}}

Jet quenching in dense matter occurs mainly at the very first stages
of the collision \cite{Gyulassy:1990ye,Baier:1994bd}. The values of the nuclear modification factor
$$R_{AA}(p_T)=\frac{\frac{dN_{AA}}{d^2p_T}}{N_{\rm coll}\frac{dN_{pp}}{d^2p_T}}$$
measured in central Au+Au collisions for $p_T>3$~GeV
fall significantly below $1$.
The dependence of the nuclear modification factor on centrality 
can be understood as due to the change of the size of the opaque medium
which modifies the mean length of the path of the jet in the fireball.
On the other hand the azimuthal asymmetry of the high-$p_T$ particles is 
believed to be a consequence of the geometric eccentricity of the medium. 
The difference of the path lengths for the jets moving ``in plane'' and ``out of plane'' 
leads to an asymmetry in the jet energy loss \cite{Gyulassy:2000gk,Shuryak:2001me,Drees:2003zh}.
The angle-averaged nuclear modification factor $R_{AA}$ 
is very weakly dependent on the
shape of the opaque medium, once its size has been fixed.

In this paper we use a simple model \cite{Drees:2003zh,Horowitz:2005ja} of the energy loss in order to explore
 the role of the shape of the event-by-event rotated absorbing medium.
Neglecting the transverse expansion of the fireball for early times relevant for jet quenching,
 we expect that the shape of the medium is close to the initial 
conditions as obtained from the distribution of sources in
the transverse plane from the Glauber models.
Several prescriptions for the distribution of the density
in the transverse 
plane  have been used, such as the wounded nucleon density, the binary collisions density, the mixed density, or the color glass condensate estimate.
All of these approaches use an event-averaged shape of  the medium in which
 the jets propagate. However, it is clear that in each particular event 
 the thermalized dense medium has a slightly different shape and position with 
respect to the geometric reaction plane.
In order to take this into account, one can use the variable-axes density
$f^\ast(\rho,\phi)$ as the density of the scattering centers for the
 propagating parton. The resulting increase of the eccentricity of the medium is expected to increase the asymmetry of the jet absorption. 
A very similar effect has been
 discussed for source the profiles calculated including saturation 
in the CGC model. Drescher et al. \cite{Drescher:2006pi} have found an increase in $v_2$
 by about $10-15\%$.

The partons are produced in p-p collisions with the power-law spectrum ${dN}/{dp_{T}^2}\propto {1}/{p_{T}^{8.1}}$ (the fragmentation is not included), and the
energy loss is taken as
\begin{equation}
\label {eloss}
\Delta E = \mu E \int_{0}^{\infty} l dl \frac{\tau_0}{l+\tau_0}
f^\ast(x_0+v_x (l+\tau_0),y_0+v_y (l+\tau_0)),
\end{equation}
where the jet production point $(x_0, \ y_0)$ is generated 
from a binary collision in the fixed-axes frame.
The fact that we must use the fixed-axes frame here results from 
an absence of correlations of the very rare jet-production collisions
and the soft collisions generating the opaque medium. 
In Eq.~(\ref{eloss}) the initial time is denoted as $\tau_0$, while the 
time measured from $\tau_0$ is $l$. 
The direction
of the parton transverse velocity $(v_x,\ v_y)$ is chosen randomly.
 For  each choice of the model of the opaque medium the parameter $\mu$ in the energy loss 
formula (\ref{eloss}) is fitted to reproduce the high-$p_T$ nuclear modification factor $R_{AA}$ \cite{Adler:2003au}.
 Then the elliptic flow coefficient 
is calculated at different centralities
($v_2$ and $R_{AA}$ are $p_{T}$-independent in such a simplified model).
The variable-axes medium has  a different shape from the fixed-axes one,
with a larger eccentricity. As has been noticed, the raise in the geometrical eccentricity increases the asymmetry of the energy loss 
\cite{Drescher:2006pi}. The variable-axes medium in the hot-spot model
has an eccentricity of about $0.4$, and  the CGC calculation gives $0.5$ at intermediate impact parameters, therefore one would expect a similar increase in $v_2$ at high $p_T$.
However, there is one important effect that should be taken into account for the event-by-event modified absorbing medium.  
The absorbing medium formed in each event is rotated and also shifted.
The shift with respect to the fixed-axes frame is quite important ({\em cf.} Fig.~\ref{fig:shifts}), yielding about 
$1/3$ of the total effect.

\begin{figure}[tb]
\begin{center}
\includegraphics[width=.42\textwidth]{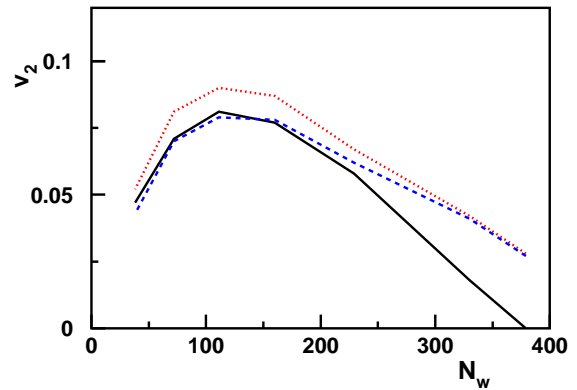} 
\end{center}
\vspace{-4mm}
\caption{(Color online) Elliptic flow coefficient at high $p_T$ as a function of the number of wounded
nucleons, obtained using the 
fixed-axes density of the wounded nucleons $f(x,y)$ in the energy loss formula \ref{eloss} (solid line), and  
with the variable-axes density $f^\ast(x,y)$ 
for the hot-spot scenario (dashed line). The dotted line represents the 
result for the variable-axes density but without the shift and rotation of the opaque medium.  
\label{fig:v2}}
\end{figure}

The resulting  elliptic flow (Fig.~\ref{fig:v2}) at centralities larger than $20\%$
resulting from the energy loss calculated with 
the wounded-nucleon model in the fixed-axes frame (solid line) comes out 
remarkably similar  to the 
result of the hot-spot model in the variable-axes frame (dashed line).
Only when the shift and rotation of the opaque medium are neglected (dotted line)
the modification of the shape leads to an increase of the high $p_T$
elliptic flow coefficient $v_2$ by about $10-15\%$, 
We have checked that the cancellation of the effects of the increased 
eccentricity of the medium and of the shift and rotation with respect
to the jet emission points at larger centralities happens also for other considered models.

In the forthcoming experiments at the Large Hadron Collider (LHC) the analysis of jet tomography with 
respect the the event-by-event reconstructed plane must take into account the relative shift and 
rotation effects discussed above.

\section{Statistical interpretation of the variable-axes moments\label{sec:moments}}

In this section we analyze the variable-axes moments and profiles from the viewpoint of statistical methods. 
The purpose of this study is to understand certain features of the numerical results presented earlier
on more general formal grounds.
It turns out that quite
simple expressions can be found for the case where correlations between the location of sources are neglected, 
which allows for the standard usage of the central limit theorem.
The Glauber models do induce some correlations, as can be seen from Fig.~\ref{fig:snap}. 
For instance, a nucleon from the skin of one of the nuclei, as present in the middle
picture, wounds several nucleons from the other nucleus. As a result, correlation between the locations of the wounded nucleons is 
generated. For simplicity, we neglect all such correlations in the analytic analysis of this section. Their role is 
discussed in Appendix~\ref{app:case2}.

If such correlations are strong, 
their analytic inclusion is difficult and one has to resort to numerical simulations such as those presented in the 
earlier sections. We also take all weights equal to unity, $w_i=1$, in order to avoid notational complications.

Appendix \ref{sec:toy} contains a very simple analysis in a one-dimensional independent-particle toy model, which avoids some notational 
complications but grasps all essential features of the full case.

From definition of the variable-axes moment, we need to evaluate 
\begin{eqnarray}
\varepsilon_{k,l}^\ast &=& \langle \langle \frac{\frac{1}{n}\sum_{j=1}^n \rho_j^k\cos[l (\phi_j-\phi^\ast)]}
{\frac{1}{n}\sum_{j=1}^n \rho_j^k} \rangle \rangle \nonumber \\
&=& \frac{1}{I_{k,0}}\langle \langle \frac{1}{n}\sum_{j=1}^n \rho_j^k\cos[l (\phi_j-\phi^\ast)] \rangle \rangle , \label{f2rgen}
\end{eqnarray}
with $\langle \langle . \rangle \rangle$ denoting the averaging over (infinitely many) events and $j$ labeling the 
source.
The weighting, according to the definition (\ref{epsilon}), is done with 
the transverse radius to the power $k$, {\em i.e.} $\rho_j^k$. 
Let us introduce the notation
\begin{eqnarray}
Y_l=\frac{1}{n}\sum_{j=1}^n \rho_j^k \cos(l \phi_j), \;\; X_l=\frac{1}{n}\sum_{j=1}^n \rho_j^k \sin(l \phi_j) \label{XY}
\end{eqnarray}
(recall that we measure the azimuthal angle from the $y$-axis).
The rotation angle $\phi^\ast$
depends on the distribution of particles in the given event, by 
definition maximizing the {\em quadrupole} moment $(l=2)$, {\em i.e.} the quantity
$\frac{1}{n}\sum_{j=1}^n \rho_j^k \cos[2 (\phi_j-\phi^\ast)]$. This gives the relations
\begin{eqnarray}
\cos(2\phi^\ast)&=&Y_2/\sqrt{Y_2^2+X_2^2}, \nonumber \\
\sin(2\phi^\ast)&=&X_2/\sqrt{Y_2^2+X_2^2}, \label{gensums}
\end{eqnarray}
Let us denote $x_j=(\rho_j,\phi_j)$ as the short-hand notation for the polar coordinates of the source point, 
and $F(x_1,x_2,\dots,x_n)$ as the $n$-particle probability distribution. Then 
we can rewrite Eq.~(\ref{f2rgen}) as 
\begin{eqnarray}
&& \varepsilon_{k,l}^\ast = \frac{1}{I_{k,0}}\int dx_1 \dots dx_n F(x_1, \dots, x_n) \times \label{f2rgen2}\\
&& \frac{1}{n}\sum_{j=1}^n \rho_j^k [ \cos(l \phi_j) \cos(2\phi^\ast)-\sin(l \phi_j) \sin(2\phi^\ast)]= \nonumber \\
&& \frac{1}{I_{k,0}}\int dx_1 \dots dx_n F(x_1, \dots, x_n) \frac{Y_l Y_2 + X_l X_2}{\sqrt{Y_2^2+X_2^2}}. \nonumber 
\end{eqnarray}

Let us analyze the quadrupole moment $\varepsilon_{k,2}^\ast$, which is the simplest but also the most 
important measure. For that case Eq.~(\ref{f2rgen}) becomes
\begin{eqnarray}
&& \varepsilon_{k,2}^\ast =
\frac{1}{I_{k,0}}\int dx_1 \dots dx_n F(x_1, \dots, x_n) \sqrt{Y_2^2+X_2^2} =  \nonumber\\
&&\frac{1}{I_{k,0}} \langle \langle   \sqrt{Y_2^2+X_2^2} \rangle \rangle = \label{case2} \\
&& \frac{1}{n I_{k,0}}\langle \langle \sqrt{\left (\sum_{j=1}^n \rho_j^k \cos(2 \phi_j) \right )^2+ 
\left (\sum_{j=1}^n \rho_j^k \sin(2 \phi_j) \right )^2} \rangle \rangle . \nonumber
\end{eqnarray}
Thus, the variable-axes quadrupole moment corresponds to a highly ``non-local'' average, involving infinitely many moments 
through the square root function. 

In the absence of correlations between collision points the many-particle probability distribution 
factorizes into $F(x_1,\dots, x_n)=f(x_1) \dots f(x_n)$, where $f(x_i)$ are normalized to unity. 
Hence all the information on the system is contained in the 
one-particle distribution functions (\ref{f}), or, equivalently, in the {\em fixed-axes} profiles $f_l(\rho)$. 
The variable-axes profiles 
$f_l^\ast(\rho)$ and moments $\varepsilon_{k,l}^\ast$ are then expressible in terms of the fixed-axes quantities. 
In order to get some useful relations, we use a method similar to the techniques 
of Refs.~\cite{Ollitrault:1997di,Poskanzer:1998yz,Borghini:2000sa}.
In the limit of large $n$ the goal is accomplished with the help of the
central limit theorem. Indeed, the variables $\rho^k \cos(2 \phi)$ and $\rho^k \sin(2 \phi)$ entering Eq.~(\ref{case2}) 
are independent from one another, since 
$\int_0^{2\pi} \rho^{2k} \cos(2 \phi) \sin(2 \phi) =0$. Therefore for sufficiently large values of $n$ one may use the 
central limit theorem, implying the normal distribution for the variables $Y_2$ and $X_2$. 
The details of this calculation are given in Appendix \ref{app:case2}.
The final results can be cast in the form of a series involving the confluent hypergeometric functions:
\begin{eqnarray}
&&\varepsilon_{k,2}^\ast = \frac{\sqrt{2} \sigma_{Y_2}^2}{I_{k,0} \sqrt{\pi } \sigma_{X_2}} \sum _{m=0}^{\infty } (2 \delta \sigma_{Y_2}^2)^m  
\times \label{epsv} \\ && \;\;\;\;\frac{
\Gamma \left(m+\frac{1}{2}\right) \Gamma \left(m+\frac{3}{2}\right) \,
   _1F_1\left(-\frac{1}{2};m+1;-\frac{\bar Y_2^2}{2 \sigma_{Y_2}^2}\right)}{ m!^2}, \nonumber 
\end{eqnarray}
where the average and standard deviation of the variables (\ref{XY}) are (see Appendix \ref{app:case2})
\begin{eqnarray}
\bar Y_2&=&I_{k,2}, \\
\sigma^2_{Y_2}&=& \frac{1}{2n}(I_{2k,0}-2 I_{k,2}^2+ I_{2k,4}) \nonumber \\
\sigma^2_{X_2}&=& \frac{1}{2n}(I_{2k,0}-I_{2k,4}), \nonumber \\
\delta &=& \frac{1}{2 \sigma_{Y_2}^2}-\frac{1}{2 \sigma_{X_2}^2}. \nonumber
\end{eqnarray}

For the special case of central collisions (where $\delta=0$) only the $m=0$ piece contributes to the series 
(\ref{epsv}) and we have
the very simple result
\begin{eqnarray}
\varepsilon_{k,l}^\ast = \frac{\sqrt{\pi I_{2k,0}} }{2I_{k,0} \sqrt{n}},\;\;\;\;\;\;(b=0). \label{vecent}
\end{eqnarray}
Similarly, for the scaled standard deviation in central collisions we obtain (see Appendix \ref{app:case2})
\begin{eqnarray}
\frac{\Delta \varepsilon_{k,l}^\ast}{\varepsilon_{k,l}^\ast}=\sqrt{\frac{4}{\pi}-1}\simeq 0.523, \;\;\;\;\;\; (b=0) \label{ratioeps}
\end{eqnarray}
Although the above result is approximate, as it has been obtained with the assumption of no two-particle correlations
between the location of sources, it 
shows an important feature present in the Glauber simulations. The value of the scaled standard deviation 
in central collisions is close to $0.5$ and is asymptotically 
independent of $n$ ({\em cf.} Figs.~\ref{fig:epsilonsa} and \ref{fig:copper}).
The behavior is also seen in the numbers given in Table~\ref{tabb} for $b=0$.

One may also derive the expression for the profile function $f^\ast_2(\rho)$ in 
the large-$n$ limit and for the case with no correlations. 
For central collisions the result is (see Appendix \ref{app:case2})
\begin{eqnarray}
f_2^\ast(\rho) \simeq \frac{1}{2}\sqrt{\frac{\pi}{n I_{2k,0}}}\,\rho^{k} f_0(\rho), \;\;\;\;\;\; (b=0) \label{cenrot2} 
\end{eqnarray}
Thus the variable-axes quadrupole profile is proportional to the monopole profile times $\rho^k$. 
Note that this behavior reflects the chosen weighting power $k$, hence in this sense is technical 
rather than physical. One verifies that Eq.~(\ref{cenrot2}) reproduces immediately Eq.~(\ref{vecent}).

\begin{figure}
\begin{center}
\subfigure{\includegraphics[width=.5\textwidth]{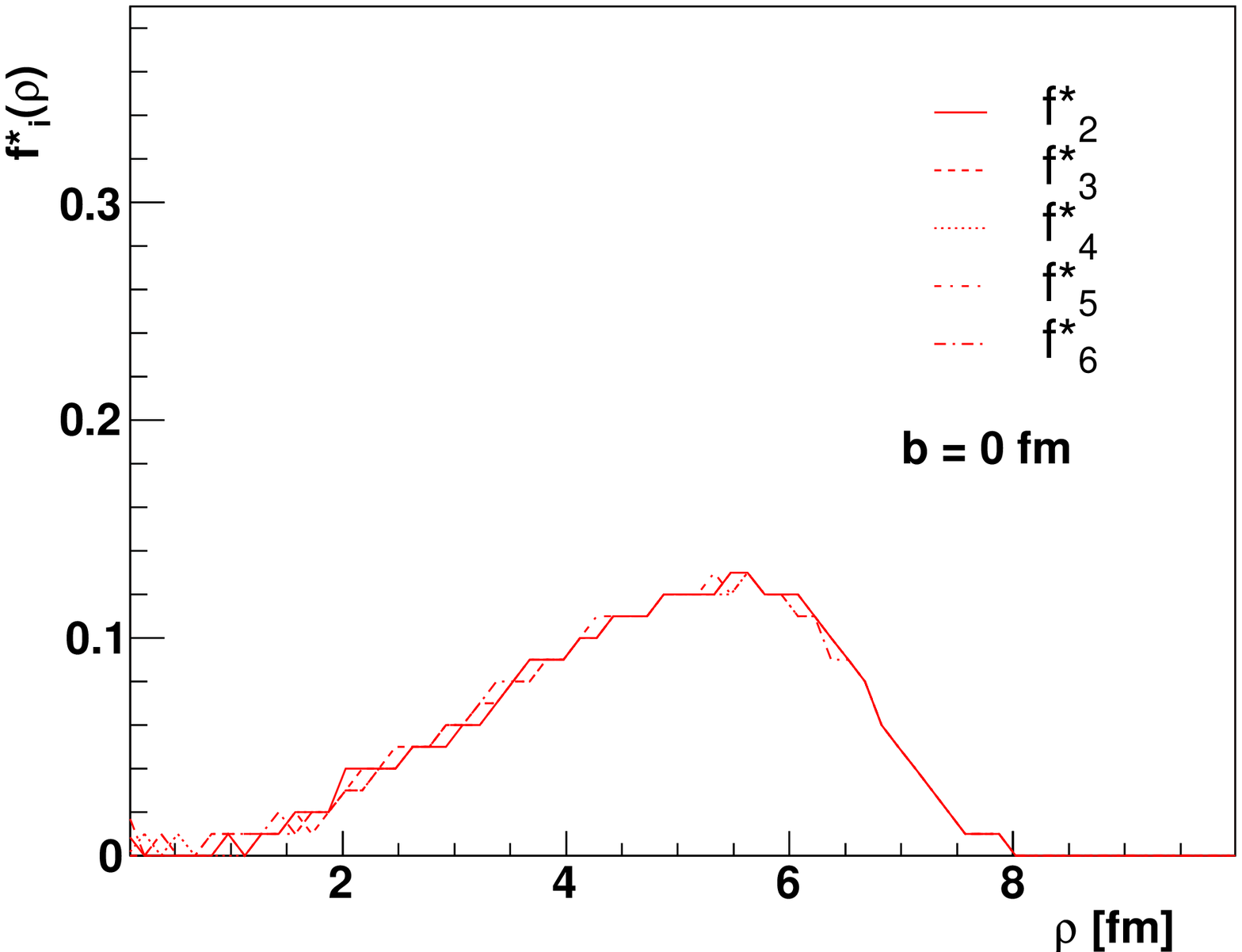}}\\ 
\subfigure{\includegraphics[width=.5\textwidth]{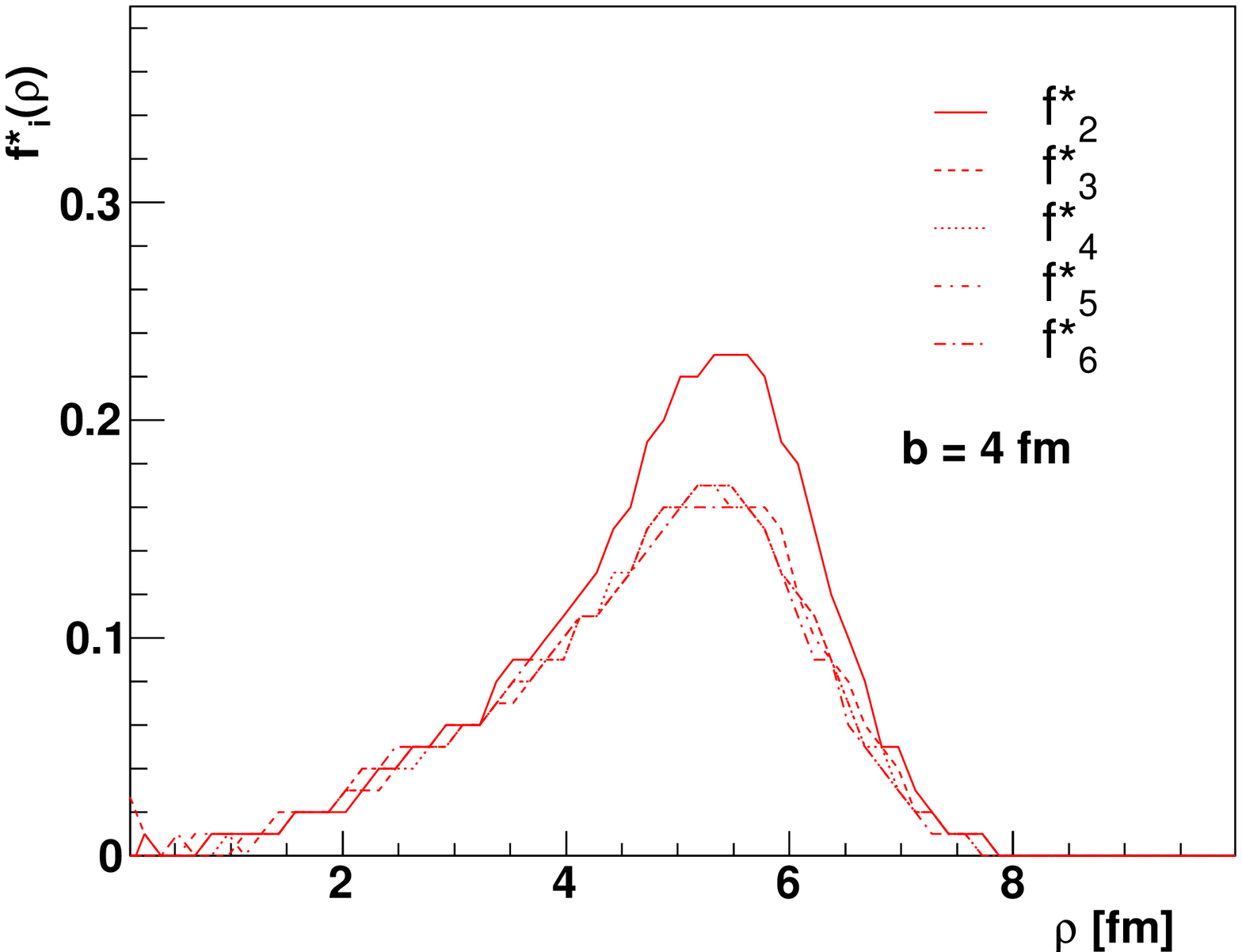}}\\
\subfigure{\includegraphics[width=.5\textwidth]{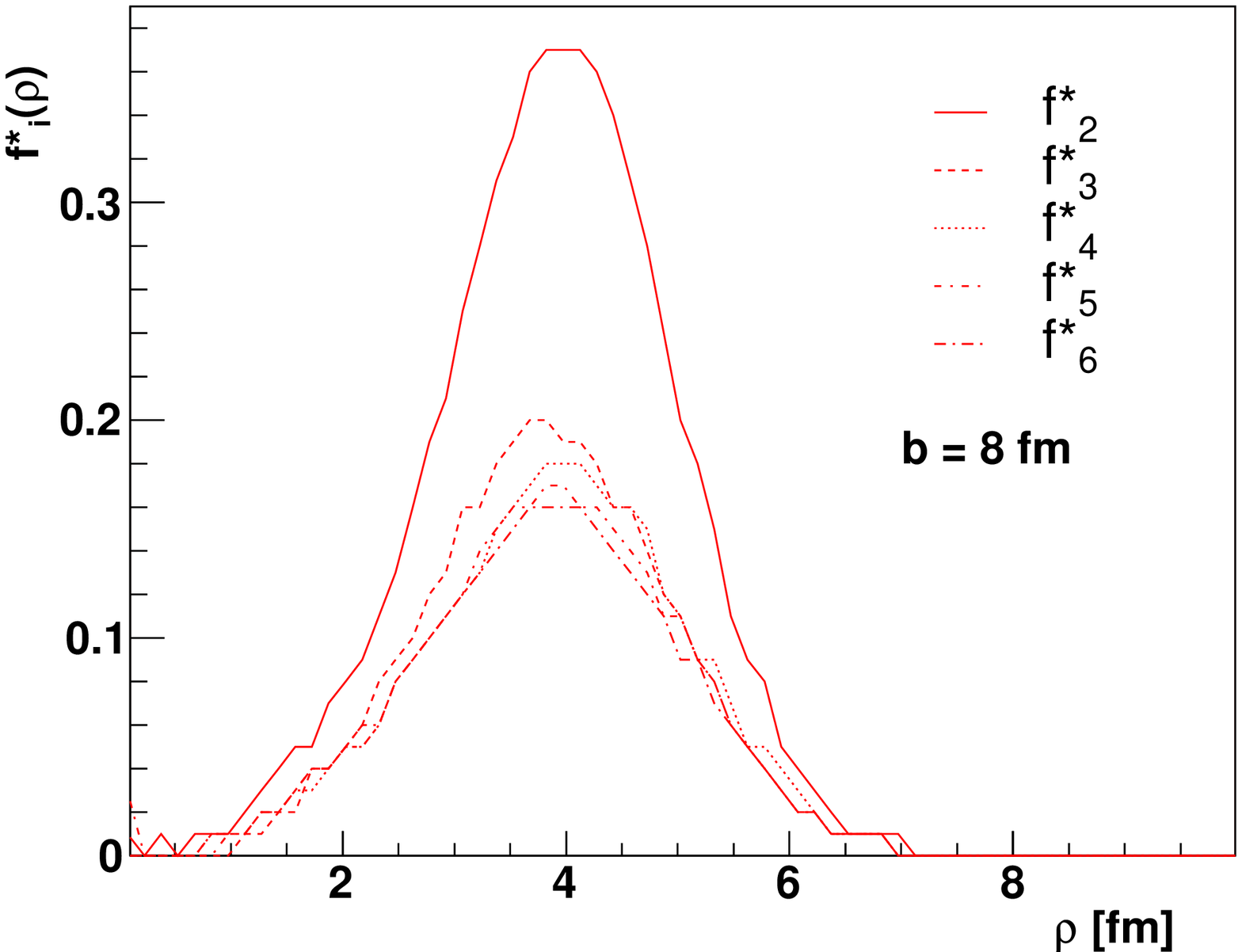}}\\
\end{center}
\caption{(Color online) The {\em multiple-axes} profiles $f^\ast_i(\rho)$, $i=2,3,4,5,6$, for the wounded nucleon model at 
several values of the impact parameter $b$. 
Gold-gold collisions. \label{fig:profmult}}
\end{figure}

\section{Multiple-axes profiles \label{sec:ee}}

In Eq.~(\ref{fr}) the rotation angle $\phi^\ast$ has been fixed with the quadrupole moment. All higher harmonics were obtained
with respect to this angle. One can give another prescription, which is superior for encoding the shape of the 
system for studies of event-by-event fluctuations. For a distribution of sources in each event we may provide the 
harmonic profile, as well as its orientation relative to the $y$ axis. Thus, each profile $f_l^\ast$ has its own 
rotation angle $\phi^\ast_l$. We introduce the expansion (the meaning of $\ast$ is now different than in 
Sect.~\ref{sec:var} and \ref{sec:flow})
\begin{eqnarray}
&&f^\ast(\rho,\phi)=f^\ast_0(\rho) + 2 f^\ast_2(\rho) \cos(2\phi-2\phi_2^\ast) \label{fre} \\ 
&& + 2 f^\ast_3(\rho) \cos(3\phi-3\phi_3^\ast)+ 2 f^\ast_4(\rho) \cos(4\phi-4\phi_4^\ast)+\dots \nonumber
\end{eqnarray}
This expansion is complete, as 
\begin{eqnarray}
\cos(l\phi-l\phi_l^\ast)&=&\cos(l\phi) \cos(l\phi_l^\ast)+\sin(l\phi) \sin(l\phi_l^\ast)=\nonumber \\
&=& a_l \cos(l\phi) +b_l \sin(l\phi),
\end{eqnarray} 
which
provides the full Fourier expansion of the distribution in each event, including both the sine and cosine
functions. Note also the presence of 
odd moments, $l=3,5,\dots$. These moments average out to zero in the limit of infinitely many events, but have non-zero 
fluctuations from event to event.  

In Fig.~\ref{fig:profmult} we show the {\em multiple-axes} profiles for the wounded nucleon model for the first few harmonics.  
We note that there is no longer a strong suppression as $l$ is increased. Clearly, to describe completely the 
full shape of the distribution we need as many harmonics as sources! 
In actual applications certain smoothing must be included, which effectively cuts off the higher harmonics from the 
expansion. These equilibration processes cause sharp shapes to smooth out, 
{\em i.e.} high harmonics are damped. One may therefore propose a smoothing prescription based on multiple-axes 
profiles, introducing a suitable cut-off function in the Fourier index $l$. Details and application to event-by-event
hydrodynamics will be presented elsewhere.

\begin{figure}
\includegraphics[width=.5\textwidth]{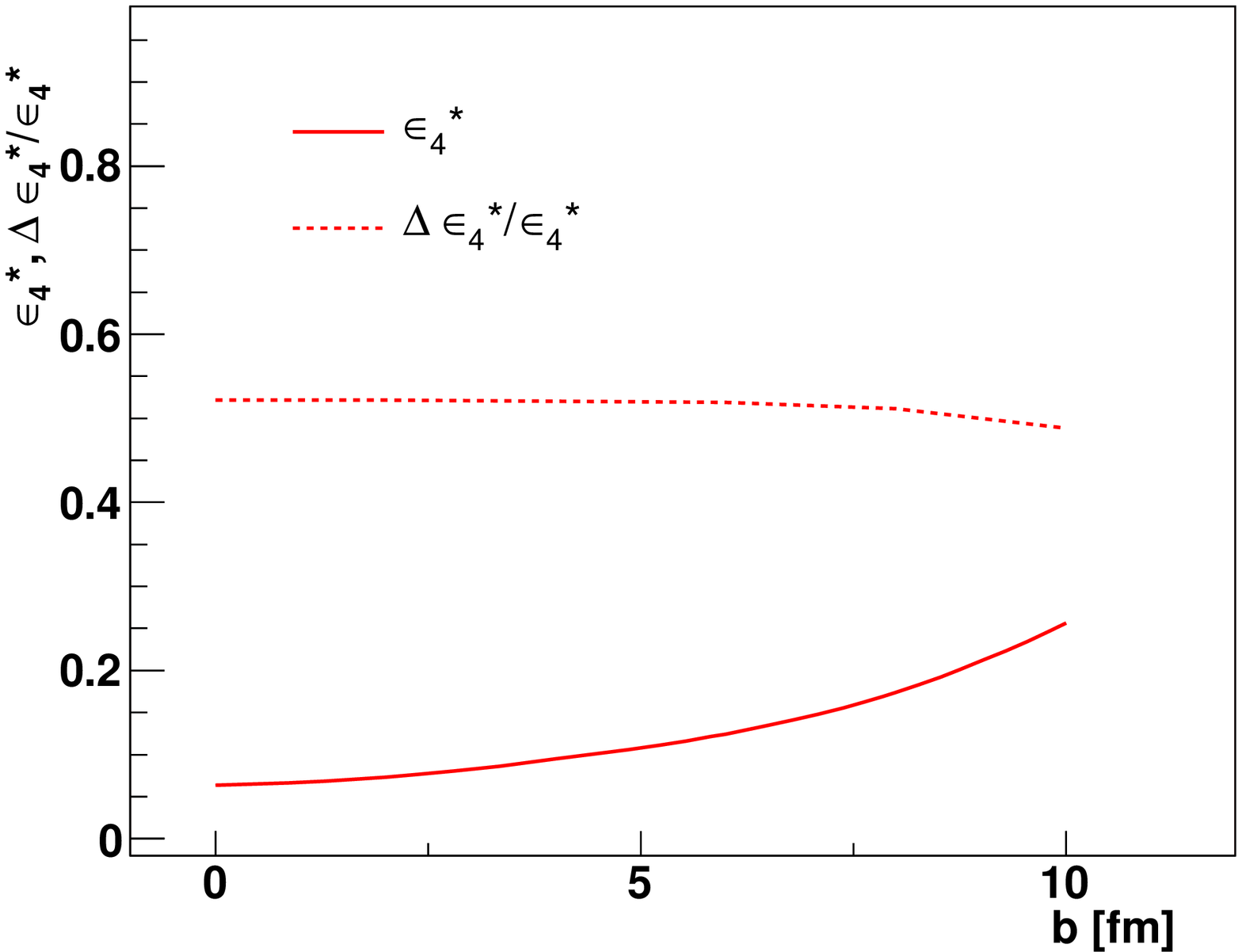}
\caption{(Color online) The multiple-axes octupole moment, $\varepsilon^\ast_4\equiv\varepsilon^\ast_{2,4}$, and 
its scaled standard deviation, plotted as functions of the impact parameter. Wounded nucleon model, gold-gold
collisions. \label{fig:eps4}}
\end{figure}

In Fig.~\ref{fig:eps4} we show the results for the multiple-axes octupole moment, $\varepsilon^\ast_{4,2}$, and 
its scaled standard deviation for the case of gold-gold collisions. 
As explained above, this moment is computed in each event in the reference frame
which maximizes the octupole distribution. We note that again at $b=0$ 
the scaled standard deviation is close to 0.5, as in 
the large-$n$ limit and in the absence of correlations it assumes the value
$\Delta \varepsilon^\ast_{l,k}/\varepsilon^\ast_{l,k}(b=0) = \sqrt{4/\pi-1}$. The details of the 
analysis are given in Appendix~\ref{app:case2}.

\section{Fluctuations of the elliptic flow\label{sec:flow}}

The statistical analysis for the fluctuations of the variable-axes shape parameters, in particular 
$\varepsilon^\ast$, carries over to the fluctuations of the elliptic flow
coefficient $v_2^\ast$. These fluctuations, which are an important 
probe of the nature of the early-stage dynamics of the system 
\cite{Mrowczynski:2002bw}, have recently been measured at RHIC
\cite{Alver:2006wh,Sorensen:2006nw,Alver:2007rm}. In fact, the experimental 
procedure used in these analyses identifies the elliptic flow coefficient with 
the {\em participant} or variable axes $v_2$, here denoted as $v_2^\ast$.

The relevance of studies of fluctuations of the initial shape of the fireball comes from the 
well-known fact that 
for small elliptic asymmetry one expects on hydrodynamic grounds the relation
\begin{eqnarray}
\frac{\Delta v_2^\ast}{v_2^\ast}=\frac{\Delta \varepsilon^\ast}{\varepsilon^\ast}. \label{hydrofl}
\end{eqnarray}
As argued in Ref.~\cite{Vogel:2007yq}, the result (\ref{hydrofl})
indicates that the mean free path in the matter created in the initial stages of the heavy-ion collisions
is very small, although turbulence does not develop. 

The statistical method used in Sect.~\ref{sec:var} carries over to $v_2^\ast$. An immediate consequence of 
Eq.~(\ref{hydrofl}), under the assumption
of the absence of correlations between the location of sources, is the result for the variable-axes coefficient, $v_2^\ast$, in central
collisions:  
\begin{eqnarray}
\frac{\Delta v_2^\ast}{v_2^\ast}(b=0)\simeq\sqrt{\frac{4}{\pi}-1} \simeq 0.52. \label{v2res}
\end{eqnarray}
The values obtained in Refs.~\cite{Sorensen:2006nw,Alver:2007rm} for 
the scaled standard deviation of $v_2^\ast$ are between 0.35 and 0.5 for all impact parameters.

An argumentation for the result (\ref{hydrofl}) may be done on general grounds as follows:
schematically, one may denote the hydrodynamic equations as $L(\psi)=0$, 
where $L$ is the operator for hydrodynamics (involving partial 
differentiation, {\em etc.}), and $\psi$ is the set of hydrodynamic 
functions of space-time describing the state of the system. If the evolution is {\em smooth}, one may expand 
to first order around the azimuthally-symmetric system $\psi_0$, 
\begin{eqnarray}
L(\psi) = L(\psi_0+\delta \psi) \simeq L(\psi_0) + L'(\psi_0)\delta \psi, \label{schem}
\end{eqnarray}
where $\delta \psi$ is the asymmetric piece, and the prime denotes the differentiation 
with respect to the hydrodynamic variables $\psi$. 
Since $L(\psi_0)=0$, we have to first order $L'(\psi_0)\delta \psi=0$.
Then, due to linearity of the
equation for $\delta \psi$, we have $||\delta \psi(t)|| \sim ||\delta \psi(t_0)||$, {\em i.e.}, 
the magnitude of the solution at time $t$ is proportional to the initial condition at $t_0$.
This concerns all hydrodynamic properties, in particular the shape and flow. As a result, 
the $v_2^\ast$ coefficient determined from the momentum spectra at time $t$ is proportional to the 
initial spatial quadrupole asymmetry $\varepsilon^\ast$. The feature holds event-by-event, 
hence the result (\ref{hydrofl}) follows. Thus Eq.~(\ref{hydrofl}) is a consequence of 
applicability of perturbation theory for the small departure from cylindrical symmetry.

We note that Eq. (\ref{hydrofl}) holds separately for the fixed-axes and the variable-axes analyses, with 
the obvious requirement to use the same method on both sides of the equation. 

One may ask if a similar argumentation can be used for the higher harmonic flow 
coefficients, $v_4^\ast$, {\em etc}. The results of the previous sections show a strong suppression of 
subsequent harmonic moments of the distribution of sources in the case of the fixed-axes and 
variable-axes analyses. This suggests the hierarchy
\begin{eqnarray}
\psi=\psi_0+\lambda \delta \psi_2 + \lambda^2 \delta \psi_4+\dots, 
\end{eqnarray} 
where $\lambda$, typically of the order of a few percent, 
is the small expansion parameter, while the subscripts $0,2,4,\dots$ label the harmonics.
Expansion of the hydrodynamic evolution to second order in $\lambda$, again under the assumption of smoothness, yields now
\begin{eqnarray}
L(\psi) &=& L(\psi_0) + \lambda L'(\psi_0)\delta \psi_2  \\&+& \label{schem2}
\lambda^2 \left [   L'(\psi_0)\delta \psi_4 +  L''(\psi_0) (\delta \psi_2)^2/2  \right ]+\dots \nonumber
\end{eqnarray}
We carry out the perturbation theory extracting the second-order equation, 
\begin{eqnarray}
L'(\psi_0)\delta \psi_4 = -  L''(\psi_0) (\delta \psi_2)^2/2, \label{2ord}
\end{eqnarray} 
and note that the evolution of $\delta \psi_4$ is coupled to $(\delta \psi_2)^2$, which acts as a source term
in the linear inhomogeneous equation for $\delta \psi_4$.

Various harmonic components evolve hydrodynamically with different 
time scales. Let us denote $\tau_2$ as the characteristic time for the operator $L'(\psi_0)$, or $\delta \psi_2$, 
and $\tau_4$ as the characteristic time for the operator $L''(\psi_0)$.
If 
\begin{eqnarray}
\tau_2 \gg \tau_4,
\end{eqnarray}
then the time scale for the source term in Eq.~(\ref{2ord}) is much larger than for the operator
$L''(\psi_0)$. In that case for $t \gg t_0$  
\begin{eqnarray}
||\psi_4(t)|| \sim ||\delta \psi_2(t)||^2 \sim ||\delta \psi_2(t_0)||^2.
\end{eqnarray}
In words, at late times the octupole deformation is proportional to the square of the initial 
quadrupole deformation, and looses memory of the initial octupole deformation $||\psi_4(t_0)||$. 
In particular, this means that
\begin{eqnarray}
v_4^\ast \sim \varepsilon^{\ast 2} \sim v_2^{\ast 2}. \label{ev2}
\end{eqnarray}
In Ref.~\cite{Borghini:2005kd} the variable $v_4/v_2^{2}$ has 
been suggested as a sensitive probe of the hydrodynamic 
evolution. Also, the simulations of Refs.~\cite{Kolb:2003zi,Borghini:2005kd} show that with increasing 
time the value of $v_2$ saturates (suggesting very large $\tau_2$), while $v_4$ quickly 
assumes the value proportional to $v_2^{2}$, supporting the assumption $\tau_2 \gg  \tau_4$ 
used in the above argumentation. 
The data of Refs.~\cite{Bai:2007ky} comply to the result (\ref{ev2}), perhaps except for very low 
values of the transverse momenta. 

For the fluctuations one gets immediately from Eq.~(\ref{ev2}) 
\begin{eqnarray}
\frac{\Delta v_4^\ast}{v_4^\ast} = 2\frac{\Delta v_2^\ast}{v_2^\ast} = 2\frac{\Delta \varepsilon^\ast}{\varepsilon^\ast}. \label{ev2fl}
\end{eqnarray}
Relation (\ref{ev2fl}), if verified experimentally, would support the scenario of smooth hydrodynamic 
evolution with the mentioned hierarchy of scales. 

At sufficiently late times {\em all} deformations are determined by the initial $\varepsilon^\ast$. 
This results in other relations, for instance
for the azimuthal Hanbury-Brown--Twiss (HBT) correlation radius, $R_{\rm HBT}(\phi)$, one expects 
\begin{eqnarray}
R_4  \sim R_2^2, \label{HBT}
\end{eqnarray} 
where $R_{\rm HBT}(\phi)=R_0+2 R_2 \cos(2 \phi)+2R_4 \cos(4 \phi) +\dots$

\section{Conclusion\label{sec:concl}}

We have presented a comprehensive study of the shape fluctuations in a variety of Glauber-like models.
Here is the list of our main points:

\begin{enumerate}

\item We compare four Glauber-like models, with different degree of fluctuation: 
the wounded-nucleon model, the mixed model, the hot-spot model, and the hot-spot model with the 
superimposed $\Gamma$ distribution. 

\item We obtain numerically the fixed-axes and variable-axes harmonic profiles and analyze their moments. 
The variable-axes moments $\varepsilon^\ast$, and 
the fixed-axes scaled standard deviation $\Delta \varepsilon/\varepsilon$ are sensitive to the choice of the variant of the Glauber model, 
while the $\Delta \varepsilon^\ast/\varepsilon^\ast$ is not, changing at most by 10-15\%. At intermediate 
values of $b$ the results of the Glauber-like models
for $\Delta \varepsilon^\ast/\varepsilon^\ast$ lie significantly above the color-glass-condensate predictions of Ref.~\cite{DNara}.  

\item We present expansions for the variable-axes moments and profiles. These analytic formulas explain the 
features of the simulations, in particular, they show that at $b=0$ the multiple-axes scaled variances are 
close to the value 0.5, insensitive of the model used, the mass number of the colliding nuclei, or the collision energy. 
In essence, the behavior of the scaled variance, used as a popular measure of the event-by-event fluctuations, is 
governed by the statistics. 
 
\item  Unlike the results of Ref.~\cite{Drescher:2006pi} which finds an increase of the jet elliptic flow $v_2$  at the level of  $10\%$, 
we find that the effect of the increased quadrupole eccentricity is largely canceled by the shift of the center 
of mass and rotation of the axes of the absorbing medium. This leads to practically no change 
of the jet emission asymmetry at intermediate and large impact parameters. Only for small impact parameters the appearance of the quadrupole moment in the shape of the medium wins over the relatively less important shift and rotation.
 
\item We propose to use an improved harmonic expansion, the multiple-axes expansion, where the harmonics in each  
event are evaluated with their own reference frame. Such a scheme may be a starting point for the event-by-event
hydrodynamic studies. The details will be presented elsewhere.

\item The analysis of the variable-axes moments in the coordinate space 
directly carries over to the collective flow and analysis of $v_2^\ast$ in the momentum space. In particular, 
Eq.~(\ref{v2res}) holds for the variable-axes elliptic flow coefficient. 

\item Finally, we comment that under plausible assumptions of smoothness, the hydrodynamic evolution leads to 
sensitivity of higher flow harmonics, $v_4$, {\em etc.} to the initial {\em quadrupole} deformation only. 
Higher harmonics of the deformation are irrelevant. Then Eq.~(\ref{ev2}) or (\ref{ev2fl}) follow.

\end{enumerate}

\acknowledgments
MR is grateful to Zbigniew W\l{}odarczyk for fruitful discussions. WB thanks Paul Sorensen, Constantin Loizides, and Wit Busza
for helpful discussions concerning the experimental determination of $v_2$ and its fluctuations. 

\appendix

\section{Details of the Monte-Carlo procedure \label{app:monte}}

The 3-dimensional positions of the nucleons in a nucleus are randomly generated from the Woods-Saxon distribution (\ref{ws}). 
Whenever the center of a nucleon is generated closer than the expulsion
distance of $d=0.4$~fm to a center of any of the prior generated nucleons, this nucleon is discarded and generated anew. 
The procedure results in a certain ``swelling'' phenomenon. 
However, for the chosen value of $d=0.4$~fm the effect is tiny, increasing the $R$ parameter by $0.01$~fm only, which is a small fraction
of a percent. The swelling effect could be compensated by reducing 
appropriately the original $R$ parameter of the distribution (\ref{ws}). At higher values of the expulsion distance 
$d$ the swelling effect increases. For instance, if one chose $d=1$~fm, then the resulting size parameter is 
$R=6.575$~fm, 3\% larger than needed, and then the compensation in the original value of $R$ should be done.

\section{Properties of the fixed-axes harmonic profiles \label{app:lowhighr}}

At fixed values of $b$ the shape of the profiles $f_l(\rho)$ reflects in a simple manner 
the average distribution of sources in a given model.
Let us assume that $d=0$, {\em i.e.} we ignore the short-range correlations, which complicate the 
analysis.  
The nucleus thickness function is 
\begin{eqnarray}
T(s)=\int_{-\infty}^\infty dz n \left ( \sqrt{s^2+z^2} \right ),
\end{eqnarray}
with the density function $n$ defined in Eq.~(\ref{ws}). The normalization is $\int 2\pi s ds T(s)=A$.
In the wounded nucleon model one has the following formula for the density of sources in the collision of 
large nuclei $A$ an $B$:
\begin{eqnarray}
n_W(\vec{b},\vec{\rho})&=&T_A(\vec{\rho}+\vec{b}/2) \left [ 1-\exp(-\sigma_{w} T_B(\vec{\rho}-\vec{b}/2) ) \right ] \nonumber \\ 
 &+& T_B(\vec{\rho}-\vec{b}/2) \left [ 1-\exp(-\sigma_{w} T_A(\vec{\rho}+\vec{b}/2) ) \right ]. \nonumber \\ \label{wn}
\end{eqnarray} 
The corresponding formula for the binary collisions is 
\begin{eqnarray}
N_{\rm bin}(\vec{b},\vec{\rho})&=&\sigma_{\rm bin} T_A(\vec{\rho}+\vec{b}/2)T_B(\vec{\rho}-\vec{b}/2). \label{bin}
\end{eqnarray} 
Let us introduce the short-hand notation $u=\rho^2+b^2/4$, $v=\rho b \sin \phi$. 
We may then rewrite Eq.~(\ref{wn},\ref{bin}) in a more explicit form 
\begin{eqnarray}
N_w(u,v )&=&T_A(u+v) \left [ 1-\exp(-\sigma_{w} T_B(u-v) ) \right ] \nonumber \\ 
 &+& T_B(u-v) \left [ 1-\exp(-\sigma_{w} T_A(u+v) ) \right ], \nonumber \\ 
N_{\rm bin}(u,v)&=& \sigma_{\rm bin} T_A(u+v)T_B(u-v). \label{wnbin2}
\end{eqnarray} 
When $\rho \ll b$ then $v \ll u$ and the low-$\rho$ expansion of the source density 
corresponding to Eqs.~(\ref{wnbin2})
has the form $\sum_n c_n \rho^{2n} 
\sin^{2n} \phi$. Because $\int_0^{2\pi} d\phi \sin^{2n}\phi \cos(2m \phi)$, needed for the
the decomposition (\ref{f}), vanishes at \mbox{$n > m$}, we obtain the result (\ref{lowr}).

The normalization constant in the Woods-Saxon function (\ref{ws}) is
\begin{eqnarray}
c=-A/\left [ 8\pi a^3 {\rm Li}_3 \left (-e^{R/a} \right) \right ].
\end{eqnarray} 
Here ${\rm Li}_n(z)=\sum_{k=1}^\infty z^k/k^n$ denotes the polylogarithm function.
At high values of $\rho$ the function (\ref{ws}) asymptotes to 
$c \exp \left ((R-r)/a) \right )$. Correspondingly, the nucleus thickness function 
at large $s$ becomes
\begin{eqnarray}
T(s) \sim c \sqrt{2\pi a s} \, e^{-s/a}.
\end{eqnarray}
For $\rho \gg b$ we also have $v \ll u$, hence with a calculation similar as for low $\rho$ we find 
Eq.~(\ref{highr}).

\section{The toy problem\label{sec:toy}}

This Appendix contains a detailed description of a toy model illustrating 
in a simple manner the statistical techniques 
used in the full-fledged calculation.
Consider the one-dimensional problem (in the azimuthal angle $\phi$) where we randomly generate uncorrelated particles from
a distribution containing the monopole and quadrupole moments only,
\begin{eqnarray}
f(\phi) = 1+2\epsilon \cos(2 \phi), \;\;\; \epsilon \in [-\frac{1}{2},\frac{1}{2}]. \label{toydist}
\end{eqnarray} 
The distribution has two fixed-axes moments, 
\begin{eqnarray}
f_0&=&\frac{1}{2\pi} \int_0^{2\pi} d\phi f(\phi)=1, \nonumber \\
f_2&=&\frac{1}{2\pi} \int_0^{2\pi} d\phi \cos(2 \phi) f(\phi)=\epsilon.
\end{eqnarray}
Suppose we generate randomly $n$ particles according to the distribution (\ref{toydist}) in each event, and 
subsequently carry the averaging of the results over the events, denoted as $\langle \langle . \rangle \rangle$.
For instance, $f_2$ is estimated as 
\begin{eqnarray}
f_2 \simeq  \langle \langle \frac{1}{n}\sum_{k=1}^n \cos(2 \phi_k) \rangle \rangle,
\end{eqnarray}
where $k$ labels the particles in each event. The equality becomes strict as the number of events approaches 
infinity, which we assume implicitly from now on.

For the variable-axes quadrupole moment we need to rotate 
the particles with the rotation angle $\phi^\ast$, which changes from event 
to event. Thus we need to evaluate ($f_2^\ast$ has the meaning of $\varepsilon^\ast$ from the other parts of the paper)
\begin{eqnarray}
f_2^\ast = \langle \langle \frac{1}{n}\sum_{k=1}^n \cos[2 (\phi_k-\phi^\ast)] \rangle \rangle. \label{toyf2r}
\end{eqnarray}
The rotation angle $\phi^\ast$ depends itself on the distribution of particles in the given event. It is such that 
the quantity $A=\frac{1}{n}\sum_{k=1}^n \cos[2 (\phi_k-\phi^\ast)]$ assumes maximum, which gives the conditions 
$dA/d\phi^\ast=0$, $d^2A/d(\phi^\ast)^2<0$. The solution is 
\begin{eqnarray}
\cos(2\phi^\ast)&=&Y_2/\sqrt{Y_2^2+X_2^2}, \nonumber \\
\sin(2\phi^\ast)&=&X_2/\sqrt{Y_2^2+X_2^2}, \label{toysums}
\end{eqnarray}
where we have introduced the short-hand notation
\begin{eqnarray}
Y_2=\frac{1}{n}\sum_{k=1}^n \cos(2 \phi_k), \;\; X_2=\frac{1}{n}\sum_{k=1}^n \sin(2 \phi_k). 
\end{eqnarray}
Using the above formulas in Eq.~(\ref{toyf2r}) yields 
\begin{eqnarray}
f_2^\ast &=&  \langle \langle \sqrt{Y_2^2+X_2^2} \rangle \rangle \label{toys} \\
&=& \langle \langle \sqrt{\left ( \frac{1}{n}\sum_{k=1}^n \cos(2 \phi_k) \right )^2+ 
\left ( \frac{1}{n}\sum_{k=1}^n \sin(2 \phi_k) \right )^2} \rangle \rangle . \nonumber
\end{eqnarray}
We see that the variable-axes moment corresponds to an average of the square root of sums (\ref{toysums}), thus is 
a highly ``non-local'' object.

For sufficiently large multiplicity of the events, $n$, one may evaluate Eq.~(\ref{toys}) with the help of the central 
limit theorem. Consider the variables $c_k = \cos (2 \phi_k)$ and \mbox{$s_k = \sin (2 \phi_k)$}. The average 
is 
\begin{eqnarray}
\bar c = \langle \langle \frac{1}{n} \sum_{k=1}^n c_k \rangle \rangle = \frac{1}{2\pi} \int_0^{2\pi}d\phi  \cos(2\phi)f(\phi)=\epsilon, 
\end{eqnarray}  
while for the variance we have 
\begin{eqnarray}
\sigma^2_c &=& \langle \langle \frac{1}{n} \sum_{k=1}^n c_k^2 \rangle \rangle -\bar c^2  \\
&=& \frac{1}{2\pi}\int_0^{2\pi}d\phi  \cos(2\phi)^2 f(\phi) - 
\epsilon^2=\frac{1}{2}-\epsilon^2. \nonumber
\end{eqnarray}  
Likewise, for the $s_k$ variable 
\begin{eqnarray}
\bar s &=& \langle \langle \frac{1}{n} \sum_{k=1}^n s_k \rangle \rangle = \frac{1}{2\pi}\int_0^{2\pi}d\phi  \sin(2\phi)f(\phi)=0, \nonumber\\
\sigma^2_s &=& \langle \langle \frac{1}{n} \sum_{k=1}^n s_k^2 \rangle \rangle = \frac{1}{2}. 
\end{eqnarray}  
Importantly, there is no correlation between $Y_2$ and $X_2$, as 
\begin{eqnarray}
\langle \langle \frac{1}{n} \sum_{k=1}^n c_k s_k \rangle \rangle = \frac{1}{2\pi}\int_0^{2\phi} d\phi \cos(2\phi) \sin(2\phi) f(\phi)=0.
\nonumber \\
\end{eqnarray}
According to the central limit theorem, the distribution of the $Y_2$ and $X_2$ variables is Gaussian,
\begin{eqnarray}
&&f(Y_2,X_2)=\frac{n}{2\pi \sigma_c \sigma_s}
\exp \left [ -n \left ( \frac{(Y_2-\bar c)^2}{2\sigma^2_c}+ \frac{X_2^2}{2\sigma^2_s}\right ) \right ] \nonumber \\
&&= \frac{n}{\pi \sqrt{1-2\epsilon^2}}
\exp \left [ -n \left ( \frac{(Y_2-\epsilon)^2}{1-2\epsilon^2}+ X_2^2 \right ) \right ].
\end{eqnarray}
Introducing the notation
\begin{eqnarray}
Y_2&=&q \cos \alpha, \;\; X_2=q \sin \alpha, \;\; q^2=Y_2^2+X_2^2, \nonumber \\
\delta&=&\frac{1}{2\sigma_c^2}-\frac{1}{2\sigma_s^2} = \frac{1}{1-2\epsilon^2} -1, \label{toydefs}
\end{eqnarray}
we may write
\begin{eqnarray}
&&f(q,\alpha)=\frac{n}{\pi \sqrt{1-2\epsilon^2}} \times  \\
&&\exp{\left [ -n \left ( \frac{q^2+\epsilon^2-2 q \epsilon \cos \alpha}{1-2\epsilon^2} \right ) + n \delta q^2 \sin^2\alpha \right ]}. \nonumber
\end{eqnarray}
What we will need  below is the integral of this distribution over $\alpha$,
\begin{eqnarray}
&&\int_0^{2\pi} d\alpha f(q,\alpha)=\frac{2n}{\sqrt{\pi} \sqrt{1-2\epsilon^2}} 
\exp{\left [ -n \left ( \frac{q^2+\epsilon^2}{1-2\epsilon^2} \right ) \right ]} \nonumber \\ && \times
\sum_{j=0}^\infty \left ( 2 q \epsilon \right )^j \frac{\Gamma(j+\frac{1}{2})}{j!}
I_j \left ( \frac{2 n \epsilon q}{1-2\epsilon^2}\right ). \label{mean}
\end{eqnarray}
As a check, it follows that 
\begin{eqnarray}
&&\int q\, dq\,d\alpha\, f(q,\alpha) \\ 
&&\;=\frac{\sqrt{1-2\epsilon^2}}{\sqrt{\pi}} \sum_{j=0}^\infty 
\left ( 2\epsilon^2 \right )^j \frac{\Gamma(j+\frac{1}{2})}{j!}=1, \nonumber
\end{eqnarray}
where we have use the definitions (\ref{toydefs}) and the formula
\begin{eqnarray}
\sum_{j=0}^\infty A^j \frac{\Gamma(j+\frac{1}{2})}{j!}=\sqrt{\frac{\pi}{1-A}}.
\end{eqnarray}

\begin{figure}[tb]
\begin{center}
\includegraphics[width=.4\textwidth]{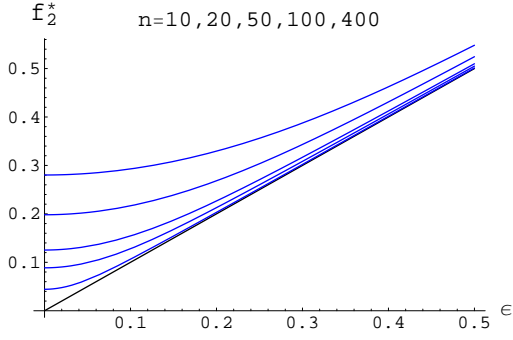} 
\end{center}
\vspace{-4mm}
\caption{(Color online) Toy model. Dependence of the variable-axes moment $f_2^\ast$ on the fixed-axes 
quadrupole moment $\epsilon$ 
for several values of the number of particles $n$. As $n$ increases, we pass from to to bottom with 
the presented curves. The straight line is the $n \to \infty$ limit, {\em i.e.} $f_2^\ast=\epsilon$. 
We note that the effect of the departure of $f_2^\ast$ from $\epsilon$ is strongest at 
low $\epsilon$ and low $n$.
\label{fig:dexp}}
\end{figure}

We may now evaluate the variable-axes moment (\ref{toyf2r}). We have the following series:
\begin{eqnarray}
f_2^\ast&=&\int q\, dq \,d\alpha \, q f(q,\alpha) = \frac{1-2\epsilon^2}{\sqrt{n \pi}} 
 \sum_{j=0}^\infty \left ( 2\epsilon^2 \right )^j \nonumber \\ &\times&
 \frac{\Gamma(j+\frac{1}{2})\Gamma(j+\frac{3}{2})}{j!^2}
{}_1F_1 \left (-\frac{1}{2},j+1;-\frac{n \epsilon^2}{1-2\epsilon^2} \right ). \nonumber \\
\label{1F1}
\end{eqnarray}
We were not able to sum up this series into a closed form, but one may readily use it 
for practical calculations in a truncated form.
At $\epsilon=0$ we have 
\begin{eqnarray}
f_2^\ast(\epsilon=0)&=&\frac{\sqrt{\pi}}{2 \sqrt{n}},
\end{eqnarray}
which shows the expected $1/\sqrt{n}$ behavior for a statistical fluctuation. 

The numerical results of the series (\ref{1F1}) are presented in Fig.~\ref{fig:dexp}.  
We note that the effect of the departure of $f_2^\ast$ from $\epsilon$ is strongest at 
low $\epsilon$ and low $n$.

\begin{figure}[tb]
\begin{center}
\includegraphics[width=.4\textwidth]{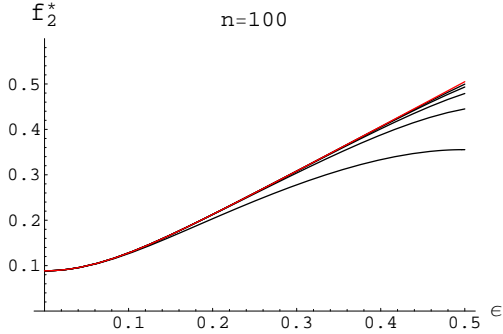} 
\end{center}
\vspace{-4mm}
\caption{(Color online) Toy model. The rate of convergence of the series (\ref{1F1}). The curves from bottom to top we show,
correspondingly, the results of summing up 1, 2, 3, 4, and 5 terms, as well as the full result.
\label{fig:dexpc}}
\end{figure}

Figure~\ref{fig:dexpc} shows the rate of convergence of the series (\ref{1F1}) for $n=100$, where we show the subsequent results of summing up 
1, 2, 3, 4, and 5 terms. We note that 5 terms are sufficient to achieve accuracy better than 1\%. 

\begin{figure}[tb]
\begin{center}
\includegraphics[width=.4\textwidth]{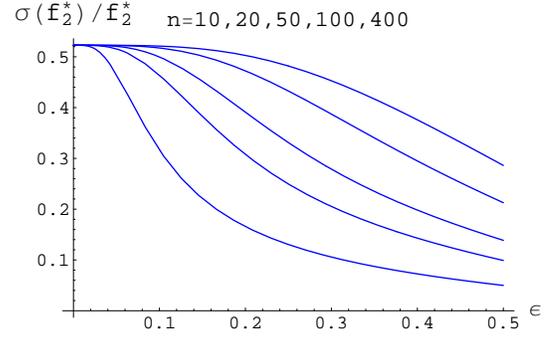} 
\end{center}
\vspace{-5mm}
\caption{(Color online) Toy model. The dependence of the scaled standard deviation on $\epsilon$ for several values of 
the number of particles $n$. 
The curves from top to bottom correspond to $n=10$, $20$, $50$, $100$, and $400$, respectively.
\label{fig:sv}}
\end{figure}

We note that an expansion of the result in powers of $\epsilon$ is useless due to slow convergence 
properties. The first few terms are
\begin{eqnarray}
f_2^\ast&=&\frac{\sqrt{\pi}}{2 \sqrt{n}}
\left (  1+\frac{n-1}{2} \epsilon^2-\frac{n^2-6 n+3}{16} \epsilon^4+{\cal O}\left( n^3 \epsilon^6\right) \right ), \nonumber \\
\end{eqnarray} 
hence the effective expansion parameter is $n \epsilon^2$. Thus for large values of $n$ the convergence radius in $\epsilon$ is very small.

The evaluation of the second moment in the $q$ variable yields
\begin{eqnarray}
&& \langle \langle Y_2^2+X_2^2 \rangle \rangle =\int q\, dq \,d\alpha \, q^2 f(q,\alpha) = \frac{\sqrt{1-2 \epsilon^2}}{n \sqrt{\pi}} 
\times \nonumber \\ && \sum_{j=0}^\infty  \frac{\left(2\epsilon^2\right)^j  \left((-2 j+n-2) \epsilon^2+j+1\right) 
\Gamma\left (j+\frac{1}{2}\right)}{j!} \nonumber \\
&& = \frac{1+(n-1) \epsilon^2}{n}.
\label{secmom} 
\end{eqnarray}
The obtained result is obvious from a direct evaluation form the definition. We compute  
\begin{eqnarray}
&&\langle \langle Y_2^2+X_2^2 \rangle \rangle = \label{toys2} \\
&&\langle \langle {\left ( \frac{1}{n}\sum_{k=1}^n \cos(2 \phi_k) \right )^2+ 
\left ( \frac{1}{n}\sum_{k=1}^n \sin(2 \phi_k) \right )^2} \rangle \rangle =\nonumber \\
&& \langle \langle \frac{1}{n}+ \frac{1}{n^2}\sum_{k \neq j} 
\left( \cos(2 \phi_j)\cos(2 \phi_k) + \sin(2 \phi_j)\sin(2 \phi_k) \right) \rangle \rangle. \nonumber
\end{eqnarray}
Since the system is uncorrelated and $\langle \cos(2 \phi_l)\rangle =\epsilon$, $\langle \sin(2 \phi_l)\rangle =0$, 
we immediately obtain 
\begin{eqnarray}
\langle \langle Y_2^2+X_2^2 \rangle \rangle = \frac{1}{n} +\frac{n(n-1) \epsilon^2}{n^2}, \label{toys3} 
\end{eqnarray}
in agreement with Eq.~(\ref{secmom}).

From Eqs.~(\ref{1F1},\ref{secmom}) we may obtain the expression for the variance of the distribution of the variable-axes moment.
A simple formula follows for the case $\epsilon=0$, where
\begin{eqnarray}
{\rm var}(f_2^\ast)=\frac{1}{n}-\left( \frac{\sqrt{\pi}}{2\sqrt{n}} \right)^2 = \frac{1-\frac{\pi}{4}}{n}\simeq \frac{0.215}{n}.
\end{eqnarray}
The scaled variance and scaled standard deviation are
\begin{eqnarray}
\frac{{\rm var}(f_2^\ast)}{f_2^\ast}&=&\frac{\frac{2}{\sqrt{\pi}}-\frac{\sqrt{\pi}}{2}}{\sqrt{n}} \simeq \frac{0.242}{\sqrt{n}}, \nonumber \\
\frac{\sigma(f_2^\ast)}{f_2^\ast}&=&\sqrt{\frac{4}{\pi}-1} \simeq 0.523. \label{estimates}
\end{eqnarray}
Note that (for $\epsilon=0$) there is no dependence on $n$ in the scaled standard deviation.  
The general case of the dependence of the scaled standard deviation of $f_2^\ast$ 
on $\epsilon$ for various values of $n$ is shown in Fig.~\ref{fig:sv}. 
The result is obtained numerically from Eq.~(\ref{1F1},\ref{secmom}).

\section{Central limit theorem and the multiple-axes moments and profiles \label{app:case2}}

This Appendix contains some details of the application of the central limit theorem to the analysis 
of the multiple-axes moments and profiles. The calculation is carried out for the case where 
each harmonic moment has its own rotation angle, as described in Sec.~\ref{sec:ee}.
Define 
\begin{eqnarray}
Y_l=\frac{1}{n}\sum_{j=1}^n \rho_j^k \cos(l \phi_j), \;\; X_l=\frac{1}{n}\sum_{j=1}^n \rho_j^k \sin(l \phi_j) \label{XYa}
\end{eqnarray}
The rotation angle $\phi^\ast$
satisfies
\begin{eqnarray}
\cos(l\phi^\ast)&=&Y_l/\sqrt{Y_l^2+X_l^2}, \nonumber \\
\sin(l\phi^\ast)&=&X_l/\sqrt{Y_l^2+X_l^2}, \label{gensumsa}
\end{eqnarray}
and it depends on the polarity index $l$. We need the averages
\begin{widetext}
\begin{eqnarray}
\bar Y_l&=&\int dx_1 \dots dx_n \frac{1}{n} \sum_{j=1}^\infty \rho_j^k \cos(l \phi_j) f(x_1,\dots,x_n) = 
\int d\phi \rho d\rho f(\rho,\phi) \rho^k \cos(l \phi) =I_{k,l}, \\
\bar X_l&=&\int dx_1 \dots dx_n \frac{1}{n} \sum_{j=1}^\infty \rho_j^k \sin(l \phi_j) f(x_1,\dots,x_n) = 
\int d\phi \rho d\rho f(\rho,\phi) \rho^k \sin(l \phi) =0, \nonumber \\
\sigma^2_{Y_l}&=& \int dx_1 \dots dx_n \frac{1}{n^2} \sum_{j=1}^\infty \rho_j^k \cos(l \phi_j)
\sum_{j'=1}^\infty \rho_{j'}^k \cos(l \phi_{j'}) f(x_1,\dots,x_n) -(\bar Y_l)^2 \nonumber \\ &&= 
\frac{1}{n}\int d\phi \rho d\rho \rho^{2k} \cos(l \phi)^2 f(\rho,\phi)+ 
\frac{n-1}{n}\left (\int d\phi \rho d\rho \rho^k \cos(l \phi) f(\rho,\phi) \right )^2-\bar q_y^2 \nonumber \\ && =
\frac{1}{2n} \int d\phi \rho d\rho \rho^{2k} f(\rho,\phi) (1+\cos(2l \phi)) - \frac{1}{n}I_{k,l}^2 =
\frac{1}{2n}(I_{2k,0}-2 I_{k,l}^2+ I_{2k,2l}) \nonumber \\
\sigma^2_{X_l}&=& \int dx_1 \dots dx_n \frac{1}{n^2} \sum_{j=1}^\infty \rho_j^k \sin(l \phi_j)
\sum_{j'=1}^\infty \rho_{j'}^k \sin(l \phi_{j'}) f(x_1,\dots,x_n) -(\bar X_l)^2 \nonumber \\ &&= 
\frac{1}{n}\int d\phi \rho d\rho \rho^{2k} \sin(l \phi)^2 f(\rho,\phi)=
\frac{1}{2n}(I_{2k,0}-I_{2k,2l}). \nonumber
\end{eqnarray}
\end{widetext}
From the central limit theorem, the distribution of $Y_l$ and $X_l$ has the normal form 
\begin{eqnarray}
f(Y_l,X_l)= \frac{1}{2\pi \sigma_{Y_l} \sigma_{X_l}} \exp \left [-\frac{(Y_l-\bar Y_l)^2}{2\sigma^2_{Y_l}}-
\frac{X_l^2}{2\sigma^2_{X_l}} \right ]. \nonumber \\
\label{normal}
\end{eqnarray}
Introducing the $q$ and $\alpha$ variables through
\begin{eqnarray}
Y_l=q \cos \alpha, \;\;\; X_l= q \sin \alpha, \label{qdef}
\end{eqnarray}
we can rewrite Eq.~(\ref{normal}) as
\begin{eqnarray}
&&f(q,\alpha) = \frac{1}{2\pi \sigma_{Y_l} \sigma_{X_l}} \times \label{normal3}\\ 
&&\;\; \exp \left [{-\frac{q^2+\bar q_y^2}{2\sigma^2_{Y_l}}+\frac{q \bar q_y \cos \alpha}{\sigma_{Y_l}^2} 
+ \delta q^2 \sin^2 \alpha} \right ], \nonumber 
\end{eqnarray}
where 
\begin{eqnarray}
\delta&=&\frac{1}{2\sigma_{Y_l}^2}-\frac{1}{2\sigma_{X_l}^2} \\ &=& \frac{2n(I_{k,l}^2-I_{2k,2l})}
{(I_{2k,0}-2I_{k,l}^2+ I_{2k,2l})(I_{2k,0}-I_{2k,2l})}.\nonumber
\end{eqnarray} 
Next, we expand in the Taylor series in $\delta$ and carry the integration over $\alpha$, which yields
\cite{Poskanzer:1998yz,Sorensen:2006nw}
\begin{eqnarray}
&& f(q) = \int_0^{2\pi} d\alpha \exp \left [ {-\frac{q^2+(\bar Y_l)^2}{2\sigma^2_{Y_l}} +
\frac{q \bar Y_l \cos \alpha}{\sigma_{Y_l}^2}} \right ]  \times \nonumber \\ 
&& \sum_{m=0}^\infty \frac{(\delta q^2 \sin^2 \alpha)^m}{2\pi \sigma_{Y_l} \sigma_{X_l} m!} 
= \exp \left [ {-\frac{q^2+\bar Y_l^2}{2 \sigma_{Y_l}^2}} \right ] \sum_{m=0}^\infty \left( \frac{2 \delta \sigma^2_{Y_l} q }{\bar Y_l} \right)^m 
\nonumber \\ && \times \frac{I_m\left({q \bar Y_l}/{\sigma_{Y_l}^2}\right) \Gamma
   \left(m+\frac{1}{2}\right)}{\sqrt{\pi }  m!\sigma_{Y_l} \sigma_{X_l}},
\end{eqnarray}
where $I_m$ is the modified Bessel function. 
The relevant moments of $f(q)$ are
\begin{eqnarray}
&&\int_0^\infty q \,dq\,f(q) = 1, \label{red} \\ 
&&\int_0^\infty q \,dq\,q f(q) = \frac{\sqrt{2} \sigma_{Y_l}^2}{\sqrt{\pi } \sigma_{X_l}} 
\sum _{m=0}^{\infty } (2 \delta \sigma_{Y_l}^2)^m  
\times \nonumber \\ && \;\;\;\;\frac{
\Gamma \left(m+\frac{1}{2}\right) \Gamma \left(m+\frac{3}{2}\right) \,
   _1F_1\left(-\frac{1}{2};m+1;-\frac{\bar Y_l^2}{2 \sigma_{Y_l}^2}\right)}{ m!^2}, \nonumber \\
&&\int_0^\infty q \,dq\,q^2 f(q) = \bar Y_l^2 +\sigma_{Y_l}^2+\sigma_{X_l}^2= \nonumber \\ && \frac{I_{2k,0}+(n-1) I_{k,l}^2}{n}, \nonumber
\end{eqnarray}
where ${}_1F_1$ is the confluent hypergeometric function.

From Eq.~(\ref{red}) one derives immediately the formulas for $l=2$ listed in the main text as Eq.~(\ref{epsv},\ref{vecent}).
For a general value of $l$ the result for the variable-axes moments (recall each moment has its own rotation angle) reads
\begin{eqnarray}
&&\varepsilon_{k,l}^\ast = \frac{\sqrt{2} \sigma_{Y_l}^2}{I_{k,0} \sqrt{\pi } \sigma_{X_l}} \sum _{m=0}^{\infty } (2 \delta \sigma_{Y_l}^2)^m  
\times \label{epsvg} \\ && \;\;\;\;\frac{
\Gamma \left(m+\frac{1}{2}\right) \Gamma \left(m+\frac{3}{2}\right) \,
   _1F_1\left(-\frac{1}{2};m+1;-\frac{\bar Y_l^2}{2 \sigma_{Y_l}^2}\right)}{ m!^2}, \nonumber \\
&&{\rm var}(\varepsilon_{k,l}^\ast)=\frac{I_{2k,0}+(n-1) I_{k,l}^2}{nI^2_{k,0}}-(\varepsilon_{k,l}^\ast)^2. \nonumber
\end{eqnarray}

\begin{figure}
\includegraphics[width=.5\textwidth]{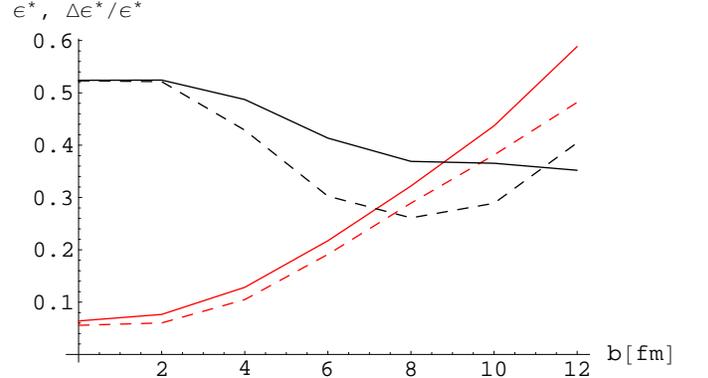}
\caption{(Color online) Comparison of the Monte Carlo calculation of $\varepsilon^\ast=\varepsilon^\ast_{2,2}$ (the rising curves) and 
$\Delta \varepsilon^\ast/\varepsilon^\ast$ in the wounded nucleon 
model for gold-gold collisions (solid lines) and the analytic formulas (\ref{epsvg}) with $l=2$ 
(dashed lines). The analytic formulas neglect
correlations between the location of sources present in the full calculation. \label{fig:Icomp2}}
\end{figure}

Figure \ref{fig:Icomp2} compares the formulas (\ref{epsvg}) for the quadrupole case ($l=2$) with the 
Monte Carlo simulation in the wounded nucleon model. The difference between the exact Monte Carlo results and the 
analytic formulas is due to the presence of correlations between the location of sources in Glauber-like models. 
These correlations result from a rather simple mechanism mentioned at the beginning of Sect.~\ref{sec:moments}: 
a nucleon from nucleus $A$ may wound several nucleons from nucleus $B$. This results in some 
clustering, hence correlations, of the locations of the wounded nucleons.
In the derivation of the analytic formulas we have resorted to the central limit theorem, hence all correlations 
were neglected. The difference between the full and uncorrelated analytic results in Fig.~\ref{fig:Icomp2} 
display the significance of the correlations.  

The correlations between the locations of sources result in a decrease of the effective number of sources $n$, so 
the full result for $\varepsilon^\ast$ (the monotonically rising curves) is naturally above the uncorrelated analytic result.
The behavior of the two curves is similar and the relative difference is at the level of 10-15\%.
For the scaled standard deviation, $\Delta \varepsilon^\ast/\varepsilon^\ast$, the comparison is more complicated. At low values of $b$ the two curves
are very close, at intermediate $b$ the calculation with correlations is higher, while at peripheric $b$ it is lower than 
the uncorrelated case. We conclude that at central collisions $\Delta \varepsilon^\ast/\varepsilon^\ast$ is not 
sensitive to correlations.   

In Monte Carlo simulations the correlations may be artificially removed by taking a very large cross section $\sigma_w$, in which 
case all the nucleons get wounded and the correlations between the locations of sources disappear. This may be
used for testing purposes. In that case the two calculations of Fig.~\ref{fig:Icomp2} overlap.

An analytic inclusion of correlations into the framework based on the central limit theorem is difficult and it is more 
productive to simply perform the simulations. However, the analytic formulas (\ref{epsvg}) bare significance not 
only at the formal level, which helps to understand the nature of the chosen statistical measures. 
There may be some models where the correlations are largely reduced compared to the wounded nucleon model, or absent. Then 
the evaluation of $\varepsilon^\ast$ and its variance are simply made by computing the moments 
$I_{2,0}$,  $I_{4,0}$, $I_{2,2}$, and $I_{4,4}$ of the {\em fixed-axes} distribution and carrying out a 
truncated series in Eq.~(\ref{epsvg}). We note that, amusingly, the 
CGC calculation of $\Delta \varepsilon^\ast/\varepsilon^\ast$ shown in Fig.~(\ref{fig:epsilonsa}) 
agrees surprisingly well with the uncorrelated result from Fig.~(\ref{fig:Icomp2}). This hints that the CGC 
approach of Ref.~\cite{DNara} has uncorrelated sources.

Next, we derive expressions for the variable-axes profiles $f_l^\ast(\rho)$ in the absence of 
particle correlations. 
These profiles correspond to inclusive distributions unintegrated over the $\rho$ variable 
of a selected particle, namely
\begin{eqnarray}
&& \!\!\!\!\!\!\!\! 2\pi \rho f_l^\ast(\rho) = 
\int d \phi \int dx_1 \dots dx_n f(x_1) \dots f(x_n) \times \nonumber \\
&& \sum_{m=1}^n \delta(\rho_m-\rho) \delta(\phi_m-\phi) \cos[l(\phi-\phi^\ast)] ,
\label{unint}
\end{eqnarray}
where the single-particle distributions $f(x_i)$ are normalized to unity.
The  inclusive distribution is normalized to $n$, hence $I_{0,0}=\int 2\pi \rho f_0^\ast(\rho)d\rho=n$.
Since all particles have equal distributions $f(x_i)$, we may relabel particles 
setting, for instance, $x_n=(\rho,\phi)$ and rewrite Eq.~(\ref{unint}) as
\begin{widetext}
\begin{eqnarray}
&& f_l^\ast(\rho) = 
\int \frac{d \phi}{2\pi} 
f(\rho,\phi) \int dx_1 \dots dx_{n-1} f(x_1) \dots f(x_{n-1}) 
\frac{1}{q}\left [ \rho^k +\cos (l \phi) \sum_{m=1}^{n-1} 
\rho_m^k \cos(l \phi_m) + \sin (l \phi) \sum_{m=1}^{n-1} \rho_m^k \sin(l \phi_m) \right ],\nonumber \\ && 
\label{unint2}
\end{eqnarray}
where we have used the definitions (\ref{gensums}) and (\ref{qdef}). 
Similarly to the analysis of the moments of Sect. \ref{sec:moments}, for sufficiently large 
values of $n$ the variables 
\begin{eqnarray}
Y_l'=\frac{1}{n}\sum_{j=1}^{n-1} \rho_j^k \cos(l \phi_j) = Y_l-\frac{1}{n} \rho^k \cos \phi = q' \cos(\alpha), \; \; 
X_l'=\frac{1}{n}\sum_{j=1}^{n-1} \rho_j^k \sin(l \phi_j) = X_l-\frac{1}{n} \rho^k \sin \phi = q' \sin(\alpha), \nonumber \\
\end{eqnarray}
follow the normal distribution with rescaled parameters 
$\bar Y_l'=\frac{n-1}{n}\bar Y_l$ and $\sigma'^2_{{Y_l},{X_l}}=\frac{n-1}{n} \sigma^2_{{Y_l},{X_l}}$. Therefore
\begin{eqnarray}
f_l^\ast(\rho) &=& 
 \int \frac{d \phi}{2\pi} f(\rho,\phi) \int  \frac{dq' d\alpha}{2\pi \sigma_{Y_l}' \sigma_{X_l}'}\frac{q'}{q} 
 \exp \left [{-\frac{{q'}^2+ \bar Y_l'{}^2}{2{\sigma'}^2_y}+\frac{q' \bar {Y_l} \cos \alpha}{{\sigma'}_{Y_l}^2} 
+ \delta' q'^2 \sin^2 \alpha} \right ]
\left [ \rho^k  +n q' \cos (l \phi -\alpha) \right ] \nonumber \\
&=& \int \frac{d \phi}{2\pi} f(\rho,\phi) \int  \frac{dq' d\alpha}{2\pi \sigma_{Y_l}' \sigma_{X_l}'}\frac{q'}
{\sqrt{q'^2+2q'\rho^k \cos(l\phi -\alpha)/n +
\rho^{2k}/n^2}} \times \nonumber \\ &&\;\;
 \exp \left [{-\frac{{q'}^2+\bar {Y_l}^2}{2{\sigma'}^2_{Y_l}}+\frac{ {q'} \bar {Y_l} \cos \alpha}{{\sigma'}_{Y_l}^2} 
+ \delta' {q'}^2 \sin^2 \alpha} \right ]
\left [ \rho^k  +n q' \cos (l \phi -\alpha) \right ] \nonumber \\
\label{unint3}
\end{eqnarray}

For the case of central collisions, where $\delta'=0$, $\bar Y_l=0$, and ${\sigma'}_{Y_l}={\sigma'}_{X_l}$, formula
(\ref{unint3}) simplifies into
\begin{eqnarray}
f_l^\ast(\rho) &=&  f_0(\rho) 
\int  \frac{dq' d\beta}{2\pi {\sigma_{Y_l}'}^2}\frac{q'}{\sqrt{q'^2+2q'\rho^k/n  \cos \beta +
\rho^{2k}/n^2}} \exp \left [{-\frac{{q'}^2}{2{\sigma'}^2_{Y_l}}} \right ]
\left [ \rho^k  +n q' \cos \beta \right ] \\ 
&=& f_0(\rho) \int  \frac{dq'}{2\pi {\sigma_{Y_l}'}^2} \exp \left [{-\frac{{q'}^2}{2{\sigma'}^2_{Y_l}}} \right ]
2 n \left[\left(1+n q/\rho^k \right) E\left(\frac{4 n q \rho ^k}{\left(\rho ^k+n q\right)^2}\right)+
          \left(1-n q/\rho^k \right) K\left(\frac{4 n q \rho^k}{\left(\rho ^k+n q\right)^2}\right)\right] \nonumber \\
          &=&f_0(\rho) \int  \frac{dq'}{2\pi {\sigma_y'}^2} \exp \left [{-\frac{{q'}^2}{2{\sigma'}^2_y}} \right ]
\pi \rho^k (1+\frac{\rho^{2k}}{8 n^2 {q'}^2}+\dots), \nonumber
\label{unint4}
\end{eqnarray}
\end{widetext}
where $E$ and $K$ denote the elliptic integrals of the second and third kind. 
Since $q'$ is of the order of ${\sigma'}_{Y_l} \sim 1/\sqrt{n}$, subsequent terms in the expansion denoted by 
$\dots$ are suppressed with powers of $n$. 
Hence, in the large-$n$ limit we may retain only the first 
term in the expansion (the unity). The same result is obtained by first expanding $q'/q$ in inverse powers of $n$ and then carrying the integration over $\beta$.  The remaining integral over $q'$ is trivial, yielding the final expression for the central case in the absence 
of correlations: 
\begin{eqnarray}
f_l^\ast(\rho) \simeq \frac{1}{2}\sqrt{\frac{\pi}{n I_{2k,0}}}\,\rho^{k} f_0(\rho), \;\;\;\;\;(b=0).\label{cenrot} 
\end{eqnarray}
Remarkably, in this case all variable-axes profiles are equal to one another and depend only on the 
monopole profile $f_0$. 

Also note, that the power in the multiplying factor $\rho^k$ simply reflects the 
(arbitrarily) chosen power for the averaging, thus 
is a matter of methodology ({\em cf.} Table~\ref{tabb}).

We now return to the non-central case of Eq.~(\ref{unint3}). We first expand the following piece in the inverse
powers, of $n$, retaining the first two:
\begin{eqnarray}
&&\frac{(\rho^k +n q' \cos(l \phi -\alpha)) q'}{\sqrt{q'^2+2q'\rho^k/n  \cos(l \phi- \alpha) +
\rho^{2k}/n^2}} \nonumber \\
&&=n q' \cos (l \phi -\alpha)+\rho^k \sin ^2(l \phi - \alpha)+ \dots
\end{eqnarray}
We may then carry the integration over $\phi$, which gives
\begin{widetext}
\begin{eqnarray}
f_l^\ast(\rho) 
&=&  \int  \frac{dq' d\alpha}{2\pi \sigma_{Y_l}' \sigma_{X_l}'}
\exp \left [{-\frac{{q'}^2+\bar {Y_l}^2}{2{\sigma'}^2_{Y_l}}+\frac{{q'} \bar Y_l \cos \alpha}{{\sigma'}_{Y_l}^2} 
+ \delta' {q'}^2 \sin^2 \alpha} \right ] \left ( n q' f_l(\rho) \cos(\alpha) + \frac{1}{2} \rho^k f_0(\rho) 
-\frac{1}{2} \rho^k f_{2l}(\rho) \cos(2\alpha) \right ) 
\nonumber \\
\label{unint5}
\end{eqnarray}
\end{widetext}
We note the presence of three fixed-axes profiles: $f_l(\rho)$, $\rho^k f_0(\rho)$, and $\rho^k f_{2l}(\rho)$. 
The integration over $\alpha$ may be done similarly to the case of the moments, via expansion in the Bessel functions and then 
carrying out the $q'$ integration. The result, involving various confluent hypergeometric functions, is rather lengthy hence we do 
not list it here.

\bibliography{eps}

\end{document}